\documentclass[12pt]{article}
\usepackage{amsfonts,amssymb,epsfig,amsmath}
\usepackage{color}

\usepackage[vcentermath]{youngtab}

\renewcommand{\baselinestretch}{1.2}

\begin{document}

\makeatletter \@addtoreset{equation}{section} \makeatother
\renewcommand{\theequation}{\thesection.\arabic{equation}}
\renewcommand{\thefootnote}{\alph{footnote}}

\begin{titlepage}

\begin{center}
\hfill {\tt SNUTP11-001}\\
\hfill {\tt arXiv:1102.4273}

\vspace{2.5cm}

{\large\bf Refined test of AdS$_4$/CFT$_3$ correspondence 
for $\mathcal{N}=2,3$ theories}

\vspace{2cm}

\renewcommand{\thefootnote}{\alph{footnote}}

{
Sangmo Cheon$^1$, Dongmin Gang$^{1,2}$, Seok Kim$^1$ and Jaemo Park$^3$}

\vspace{1cm}

\textit{$^1$Department of Physics and Astronomy \& Center for
Theoretical Physics,\\
Seoul National University, Seoul 151-747, Korea.}\\

\vspace{0.2cm}

\textit{$^2$ Korea Institute for Advanced Study,
 Seoul 130-012, Korea.}\\

\vspace{0.2 cm}

\textit{ $^3$Department of Physics \& Center for Theoretical Physics (PCTP),\\
POSTECH, Pohang 790-784, Korea.}\\

\vspace{0.7cm}


\end{center}

\vspace{2cm}

\begin{abstract}

We investigate the superconformal indices for the
Chern-Simons-matter theories proposed for M2-branes probing the
cones over $N^{010}/\mathbb{Z}_k$, $Q^{111}$, $M^{32}$ with
$\mathcal{N}=2,3$ supersymmetries and compare them with the
corresponding dual gravity indices. For $N^{010}$, we find perfect
agreements. In addition, for $N^{010}/\mathbb{Z}_k$, we also find an
agreement with the gravity index including the contributions from
two types of D6-branes wrapping $\mathbb{RP}^3$. For $Q^{111}$, we
find that the model obtained by adding fundamental flavors to the
$\mathcal{N}\!=\!6$ theory has the right structure to be the correct
model. For $M^{32}$, we find the matching with the
gravity index modulo contributions from peculiar saddle points.
\end{abstract}

\end{titlepage}

\renewcommand{\thefootnote}{\arabic{footnote}}

\setcounter{footnote}{0}

\renewcommand{\baselinestretch}{1}

\tableofcontents

\renewcommand{\baselinestretch}{1.2}

\section{Introduction}

Recent years have seen the tremendous development in our
understanding of $AdS_4/CFT_3$ correspondence. Important inflection
point is proposal by J. Schwarz \cite{sch} that underlying $CFT_3$
can be written as Chern-Simons matter (CSM) theories without the
usual kinetic term for the gauge fields, especially for higher
supersymmetric theories with $N\ge4$. Once that proposal is realized
in specific examples, the understanding of $AdS_4/CFT_3$
correspondence has grown by leaps and bounds. After the realization
that the Bagger-Lambert-Gustavsson theory \cite{BL1,
BL2, BL3, gus1, gus2} can be written as the usual $SU(2)\times
SU(2)$ Chern-Simons matter theory \cite{rams}, there appeared a paper by Gaiotto
and Witten \cite{GaiottoWitten} where the attempt is made to write
down $N=4$ Chern-Simons matter theory with matter hypermultiplets.
The attempt was generalized in \cite{HLLLP} which includes twisted
hypermultiplets as well, thereby writing down the general classes of
$N=4$ Chern-Simons matter theories. The special case of such
construction is the famous $N=6$ theory, known as ABJM theory
\cite{ABJM, HLLLP2}, describing coincident M2 branes on $C^4/Z_k$
where $(k,-k)$ is the Chern-Simons level for two gauge groups of
ABJM theory.

Since then, various tests and checks are made for $AdS4/CFT3$
correspondence with the most emphasis on ABJM theory. The first
check is the matching of the moduli space\cite{ABJM} or chiral
operators. Also the partition function on $S^3$ of ABJM theory is
worked out to confirm the famous $N^{\frac{3}{2}}$ behavior of the
membrane degrees of freedom\cite{Drukker}. Similar behvior is also
observed for other $N=4$ theories in \cite{Herzog:2010hf}. There is
a huge development on the  integrability of the special sectors of
AdS4/CFT3 correpondence. See \cite{Integra} and subsequent reviews.

One of the most sophisticated test so far comes from the computation
and comparison of the field theory/gravity index. For ABJM theory,
such computation was first worked out for large $k$ limit in
\cite{Bhattacharya:2008bja} and was generalized for arbitrary $k$ in
\cite{Kim:2009wb} and especially for $k=1$. The computation was generalized
to $N=4$ theories in \cite{Imamura:2009hc}. The problem of the
states arising in the gravity index from the twisted sector was
resolved by \cite{Imamura:2010sa} and again the field theory index
perfectly matches with that of the gravity.

The index computation with the lower supersymmetric theories poses
several problems. First of all, given the gravity background of the
type $AdS^4 \times SE_7$ where $SE_7$ denotes a suitable
Sasaki-Einstein 7-manifold, it's not clear in general what kind of
field theory we have to compare with. There are plethora of examples
which lead to the same moduli space in field theory. One might think
that such theories are related to each other via dualities a la
Seiberg. This could be true of some related theories but there are
also counterexamples. In \cite{Kim:2010vwa}, it was shown that two
theories, ABJM theory and $N=3$ variant of the dual ABJM model lead
to the differnt indices even though they have the identical moduli
space $C^4$ with Chern-Simons level $k=1$. Going down to $N=2$
theories we have far more theories having identical chiral rings. On
the other hand in \cite{Kapustin}, it was shown that the partition
function of the theories, which are thought be related  by
Seiberg-like dualities, is the same. Thus the situation is far more
subtle.

Secondly, if we consider general $N=2$ theories, new subtleties
arise since the conformal dimension and the corresponding $R$-charge
of the fields, related by superconformal symmetry, can take
nonconventional values in IR. Until recently, it's not clear how to
tackle this problem. In \cite{Jafferis2}, Jafferis proposes that the
partition function on $S^3$ for a suitable superconformal field
theory as a function of trial $R$ charges is extremized on the actual
value of the R-charges of the underlying fields. Along with this
proposal, he writes down field theory action on $S^3$ with arbitrary
$R$ charge of the underlying fields with the symmetry $Osp(2|2)$. See also
\cite{Hama:2010av}. Similar calculation was applied to an index computation in
\cite{Imamura:2011su}. Certainly it's worthwhile to
apply these proposals in a specific example and to see if this gives
rise to the desired check for a given AdS4/CFT3 dual pair.

With the subtleties mentioned above, the goal of the paper is
to work out index for the gravity/field theory pair with $N=2$ or
$N=3$ SUSY. The first case study is done for the proposed theory for
M2 branes on $N^{010}/Z_k$, which has $N=3$ supersymmetry. The
proposed CSM theory is given by ABJM theory with flavors
\cite{Hohenegger:2009as,Gaiotto:2009tk,Hikida:2009tp}. See also \cite{Fujita:2009xz}
for generalizations. Since ABJM theory
has the gauge group $U(N) \times U(N)$ one can have $m_1$ flavors for the first
factor and $m_2$ flavors for the 2nd factor. In order to have moduli
space $N^{010}/Z_k$ we have the relation $m_1+m_2=k$ where $(k,-k)$
are the Chern-Simons level for the gauge groups. In the Type IIA
picture this theory is dual to type IIA string theory on $AdS_4\times\mathbb{CP}^3$
with D6-branes. At $k\!=\!1$, the `single D6-brane' is geometrized to $N^{010}$.
Happily, we find the perfect matching of field theory/gravity index.
Furthemore, the index captures nicely the picture of $AdS_4 \times
CP^3$ with D6 branes at general $k$ so that additional contributions
to the index can be ascribed to the string modes between different D6 branes.
In terms of the index computation, $N=3$ field theory has less
subtleties since we have firm control of F-terms and D-terms. This
case study can be contrasted with the another case study of $N=2$
theory.

The next case studies concern on specific $N=2$ theories. As we
mentioned, given an M-theory background on $AdS_4\times SE_7$,
there's no known algorithm to tell you which CSM theories we have to
consider. At this stage, one should rely on trial and errors so that
we will scan through the various theories which give rise to the
desired moduli space.

For the case of $C(Q^{111})$, there are many models suggested to
describe M2-branes probing this background. We find that the best
behaved model to be dual to the gravity on $AdS_4\times Q^{111}$ is
the one obtained by adding four fundamental flavors to the
$\mathcal{N}\!=\!6$ Chern-Simons theory \cite{Jafferis:2009th,
Benini}.\footnote{S.K. thanks D. Jafferis, I. Klebanov, S. Pufu and
B. Safdi for emphasizing the importance of this model, especially related
to their studies on the partition function on $S^3$ \cite{Jafferis:2011zi}.}
During the comparison of the gauge theory and gravity
indices, we realized that the gravity spectrum on $AdS_4\times
Q^{111}$ obtained in \cite{Merlatti:2000ed} has been cross-checked
in a very limited subsector in the literature. Among all the short
$OSp(2|4)$ multiplets listed in \cite{Merlatti:2000ed}, we find that
the large $N$ field theory index agrees with the gravity index only
after getting rid of a few towers of multiplets (which have not been
cross-checked in the literature, as far as we are aware of). To
conclusively check the validity of this model, one might have to
carefully re-examine the results of \cite{Merlatti:2000ed}. One may
also recall the model constructed in \cite{Franco:2008um} for
$Q^{111}$. We only made a preliminary study of the index of this
model, which has as many complications as our next example $M^{32}$
has. A thorough study has not been made for the models of
\cite{Franco:2008um}, but it could be that a study similar to our
section 4 may reveal a similar structure. At least, we have checked
that the index for the latter model disagrees with
\cite{Merlatti:2000ed} if we keep all multiplets claimed there.

We also find that in the former model of \cite{Jafferis:2009th,Benini}, the trial R-charge
does not appear in the large $N$ low energy spectrum (as far as the index can see).

For the case of $M^{32}$ models proposed in \cite{Martelli:2008si,Hanany:2008cd},
we find two clearly distinguished contributions to the index. The first part, coming
from a set of saddle points of the localization calculation, does not depend on
the trial R-charge and completely reproduces the gravity index on
$AdS_4\times M^{32}$. The second part, coming from the remaining
saddle points, does depend on the trial R-charge and does not seem
to correspond to any states on the gravity side. This may suggest that this model
could not be correctly describing the desired gravity dual. However, see section 4
for the possible subtleties of the calculation and also for the possibility to
construct a variant model to cure this discrepancy.

The content of the paper is as follows; After the introduction, in
section 2, we work out the index for field theory/gravity pair for
$N^{010}/Z_k$ and find perfect matching. In section 3 and 4, we
carry out similar computation for $Q^{111}$ and $M^{32}$,
respectively. In conclusion we enumerate the various future
directions.

As this work is completed, we received the paper by Imamura et al
\cite{Imamura:2011uj}, where similar topic is covered. However the
comparison with gravity side is lacking in their paper.

\section{The index for M2-branes probing $C(N^{010}/\mathbb{Z}_k)$}

\subsection{Field theory}

The field theory dual proposed for M-theory on $AdS_4\times N^{010}/\mathbb{Z}_k$
is given by the $U(N)_k\times U(N)_{-k}$ Chern-Simons theory coupled to two
bifundamental hypermultiplets and $m_1$, $m_2\!=\!k\!-\!m_1$ fundamental hypermultiplets
in the first and second gauge group, respectively. As we shall review shortly below,
the division of $k$ fundamental hypermultiplets into $m_1$ and $m_2$ has to do with
the presence of two types of D6-branes wrapping $AdS_4\times\mathbb{RP}^3$ in the gravity
dual \cite{Hikida:2009tp}, with different $\mathbb{Z}_2$ valued Wilson lines on the worldvolume.
The matter fields and their gauge and global charges are listed in Table
\ref{charges-n010}.\footnote{In \cite{Hohenegger:2009as,Imamura:2011uj}, the definition of
$U(1)_B$ is different from ours by suitably mixing with diagonal $U(1)\subset U(N)\times U(N)$
and $U(1)^2\subset U(m_1)\times U(m_2)$ flavor symmetries. Our $U(1)_B$ is the field theory
dual of the KK momentum along the M-theory circle, which is an integer multiple of $k$.}
\begin{table}[t!]
$$
\begin{array}{c|c|cc|cc|ccc}
  \hline{\rm fields}&U(N)\times U(N)&I_{SU(2)_F}&U(1)_B&U(m_1)&U(m_2)&j_3&\epsilon\\
  \hline A_{1,2}&(N,\bar{N})&\pm\frac{1}{2}&1&1&1&0&\frac{1}{2}\\
  B_{1,2}&(\bar{N},N)&\pm\frac{1}{2}&-1&1&1&0&\frac{1}{2}\\
  \hline q_i&(N,1)&0&\frac{1}{2}&m_1&1&0&\frac{1}{2}\\
  \tilde{q}_i&(\bar{N},1)&0&-\frac{1}{2}&\bar{m}_1&1&0&\frac{1}{2}\\
  \hline Q_I&(1,N)&0&-\frac{1}{2}&1&m_2&0&\frac{1}{2}\\
  \tilde{Q}_I&(1,\bar{N})&0&\frac{1}{2}&1&\bar{m}_2&0&\frac{1}{2}\\
  \hline
\end{array}
$$
\caption{charges of bosonic fields in the CFT dual of $N^{010}$}\label{charges-n010}
\end{table}
$h_0$ denotes the Cartan of $SU(2)_R$ symmetry.
In the table, $U(1)_B$ is the so-called `bayron-like' $U(1)$ charge, named after
similar symmetry of the $\mathcal{N}\!=\!6$ theory. This is expected to combine
with $SU(2)_F$ flavor symmetry to provide the enhanced $SU(3)_F$ at $k\!=\!1$.
Thus, it should not be confused with the real baryon symmetry, under which M5-branes
wrapping topological 5-cycles are charged.

The R-charge of this theory is $SU(2)_R$, so that its Cartan $h_0$ to be used to
define the BPS sector and index takes the canonical value $\pm\frac{1}{2}$ for all fields.

The superconformal index of this theory can be computed by a procedure similar to
\cite{Kim:2009wb}. The index is defined by
\begin{equation}
  {\rm Tr}\left[(-1)^Fx^{\epsilon+j_3}y_1^{I_F}y_2^B\right]\ ,
\end{equation}
where the trace is taken over the space of local gauge invariant operators, and $B$
denotes the $U(1)_B$ charge. One may also
insert chemical potentials for the $U(m_1)\times U(m_2)$ flavor symmetry, after which one
would obtain factors of fundamental/anti-fundamental characters in the letter index
explained below.

The index is given in terms of the so-called letter indices, which are given by
\begin{eqnarray}
  f^\pm&=&\frac{x^{1/2}}{1+x}(y_1^{1/2}\!+\!y_1^{-1/2})y_2^{\pm 1}\ \ \
  ({\rm from}\ B_a^\dag,\psi_{A_a+}\ {\rm or}\ A_a^\dag,\psi_{B_a+})\nonumber\\
  f^{\pm}_{1}&=&\frac{x^{1/2}}{1+x}y_2^{\pm 1/2}\ ,\ \
  f^{\pm}_{2}=\frac{x^{1/2}}{1+x}y_2^{\mp 1/2}\ .
\end{eqnarray}
The index is given by
\begin{equation}
  I(x,y_1,y_2)=\sum_{\{n_i\},\{\tilde{n}_i\}}I_{\{n_i\},\{\tilde{n}_i\}}(x,y_1 , y_2)\ ,
\end{equation}
where
\begin{align}
&I_{\{n_i\},\{\tilde{n}_i\}}(x,y_1 , y_2) \nonumber
\\
&= x^{ \epsilon_0 } \int \frac{1}{\textrm{(symmetry)}} [\frac{d\alpha_i d \tilde{\alpha}_i}{(2 \pi)^2}] e^{i  k \sum_{i=1}^N(n_ i \alpha_i - \tilde{n}_i \tilde{\alpha}_i)}  \exp [- \sum_{i\neq j} \sum_{p=1}^\infty \frac{1}{p}(x^{p|n_i - n_j|}e^{-ip(\alpha_i - \alpha_j)} + x^{p|\tilde{n}_i - \tilde{n}_j|}e^{-ip(\tilde{\alpha}_i - \tilde{\alpha}_j)})] \nonumber
\\
&\times \exp [\sum_{i,j=1}^N \sum_{p=1}^\infty \frac{1}{p}  (f^{+}(x^p, y_1^p, y_2^p) x^{p|n_i -\tilde{n}_j|} e^{ip (\tilde{\alpha}_j - \alpha_i)} + f^{-} (x^p, y_1^p, y_2^p) x^{p|n_i -\tilde{n}_j|}e^{- ip (\tilde{\alpha}_j - \alpha_i)} )] \nonumber
\\
&\times \exp [\sum_{i=1}^N \sum_{p=1}^\infty \frac{1}{p} (m_1f^+_1(x^p , y_1^p, y_2^p)x^{p|n_i|}
e^{-ip\alpha_i}\!+\!m_1 f^-_1(x^p , y_1^p, y_2^p) x^{p|n_i|} e^{ ip \alpha_i  }) + (m_1,\alpha,n,f^\pm_1 \rightarrow m_2,\tilde{\alpha},\tilde{n},f^\pm_2) ].
\label{field full index}
\end{align}
Here the summation is over all integral magnetic monopole charges, $\{n_i, \tilde{n}_i\}$.
The zero point energy $\epsilon_0$ is given by
\begin{align}
&\epsilon_0 =\sum_{i,j=1}^N |n_i - \tilde{n}_j|-\sum_{i<j}|n_i - n_j|- \sum_{i<j}|\tilde{n}_i - \tilde{n}_j| + \frac{m_1}{2} \sum_{i=1}^N |n_i| +  \frac{m_2}{2} \sum_{i=1}^N |\tilde{n}_i|.
\label{def of f, epsilon0}
\end{align}
Symmetry factor in the expression is the dimension of the Weyl group for the gauge group unbroken
by the magnetic flux \cite{Kim:2009wb}.

In the large $N$ limit, the integral over the holonomies $\alpha,\tilde\alpha$ which do not
host magnetic flux can be done by Gaussian approximation, introducing the distribution functions
$\rho(\alpha)=\sum_{n=-\infty}^\infty\rho_ne^{-in\alpha}$ and $\chi(\alpha)$.
After performing the integral, similar to \cite{Kim:2009wb}, the index takes the form of
\begin{equation}
  I_{N=\infty} (x,y_1 , y_2) = I^{(0)}(x,y_1) I'(x,y_1,y_2),\nonumber
\end{equation}
where
\begin{align}
 &I^{(0)}(x,y_1)\nonumber
\\
 & =  \prod_{n=1}^{\infty} \frac{1}{1-f^+(\cdot^n)f^-(\cdot^n)}
 \times \exp[\sum_{n=1}^\infty \frac{1}{n} \frac{\big{(}m_1^2 + m_2^2 + 2m_1m_2
 \frac{x^{n/2}}{1+x^n}(y_1^{n/2}\!+\!y_1^{-n/2})\big{)}}{1-f^+ f^-(\cdot^n)}\left(\frac{x^{n/2}}{1+x^n}\right)^2 ] \nonumber
\\
 & =  \prod_{n=1}^{\infty} \frac{(1-x^{2n})^2}{(1-x^n y_1 ^n)(1- x^n y_1^{-n})(1-x^n)^2}
 \times  \exp[\sum_{n=1}^\infty \frac{1}{n} \frac{\big{(}m_1^2 + m_2^2 + 2m_1 m_2
 \frac{x^{n/2}}{1+x^n}(y_1^{n/2}\!+\!y_1^{-n/2})\big{)}}
 {1-f^+ f^-(\cdot^n)} \left(\frac{x^{n/2}}{1+x^n}\right)^2  ]  \nonumber
\\
 &\equiv\exp [\sum_{n=1}^\infty \frac{1}n I^{(0)}_{single} (x^n, y_1^n)],
\end{align}
with
\begin{align}
 &I^{(0)}_{single} (x, y_1) = \frac{x }{y_1-x}+\frac{x y_1}{1-x y_1}+\frac{2 x}{1-x} -\frac{2 x^2}{1-x^2} \nonumber\\
 &+(m_1^2 + m_2^2) \frac{ x}{\left(1-x y_1\right)\left(1-xy_1^{-1}\right)}+2m_1m_2
 \frac{x^{3/2}(y_1^{1/2}+y_1^{-1/2})}{(1+x)\left(1-x y_1\right)\left(1-xy_1^{-1}\right) }
 \ .\label{field I(0)}
\end{align}
This part does not refer to magnetic monopole flux, and thus only contains states
which are $U(1)_B$ neutral. The remaining part $I^\prime$ is given by
\begin{align}
&I'(x,y_1, y_2) = x^{\epsilon_0}\int \frac{1}{(\textrm{symmetry})} [\frac{d\alpha}{2\pi}][\frac{d\tilde{\alpha}}{2\pi}] e^{ik \sum (n_i\alpha_i - \tilde{n}_i\tilde{\alpha}_i)}  \nonumber
\\
&\times \exp \big{[}\sum_{i=1}^{M_1}\sum_{j=1}^{M_2}  \frac{1}n f^{\rm{bif}}_{ij}(\cdot^n) + \sum_{i,j=1}^{M_1} \frac{1}{n}f^{\rm{adj}}_{ij} (\cdot^n)+ \sum_{i,j=1}^{M_2} \frac{1}{n}\tilde{f}^{\rm{adj}}_{ij} (\cdot^n)\big{]}\label{formula for I'}
\end{align}
is an integral over the holonomies associated with nonzero flux, with
\begin{align}
&f^{\rm{bif}}_{ij} = (x^{|n_i - \tilde{n}_j|} - x^{|n_i|+|\tilde{n}_j|}) (f^+ e^{i (\tilde{\alpha}_j - \alpha_i)}+f^- e^{i(\alpha_i -\tilde{\alpha}_j)}), \nonumber
\\
&f^{\rm{adj}}_{ij} = -\big{[}(1-\delta_{ij})x^{|n_i - n_j|} - x^{|n_i| + |n_j| } \big{]}e^{- i (\alpha_i -\alpha_j)}, \nonumber
\\
&\tilde{f}^{\rm{adj}}_{ij} = -\big{[}(1-\delta_{ij})x^{|\tilde{n}_i - \tilde{n}_j|} - x^{|\tilde{n}_i| + |\tilde{n}_j| } \big{]}e^{- i (\tilde{\alpha}_i -\tilde{\alpha}_j)}.
\end{align}
$M_1$ and $M_2$ are number of nonzero fluxes.
Like \cite{Kim:2009wb}, $I'(x,y_1, y_2)$ can be factorized and yields
\begin{align}
I_{N=\infty} (x,y_1, y_2)=I^{(0)} (x, y_1) I^{(+)} (x,y_1, y_2) I^{(-)} (x,y_1, y_2).
\end{align}
$I^{\pm}$ can be computed using the formula for $I'$ in \eqref{formula for I'} with all monopole
charges are positive/negative. $I^\pm$ can be expanded in positive/negative powers of $y_2$,
respectively.

Let us leave some remarks on the structure of the large $N$ index.
Firstly, the `single particle index' $I^{(0)}_{single}$ in (\ref{field I(0)}) consists
of two parts. As will be explained later, the first line is identical to the single
graviton index on $AdS_4\times\mathbb{CP}^3$. The second line will be shown to be the
single particle index of the open string degrees of freedom living on $m_1$ and $m_2$
D6-branes with different $\mathbb{Z}_2$ Wilson lines, wrapping $AdS_4\times\mathbb{RP}^3$
in $AdS_4\times\mathbb{CP}^3$. See the next subsection for more explanations.

Secondly, in the index $I^\pm$ with monopoles, the fundamental degrees of freedom
encoded by the indices $f^\pm_1$ and $f^\pm_2$ totally disappears, and the only trace of
their existence is in the zero point energy $\epsilon_0$. The gravity dual interpretation
of this phenomena would be the absence of open string modes between D0-D6 branes of type IIA
theory on $AdS_4\times\mathbb{CP}^3$.
Also, apart from the last zero point energy factor $x^{\frac{1}{2}\sum(m_1|n_i|+m_2|\tilde{n}_i|)}$,
the integral is exactly the same as the $I^+$ of $\mathcal{N}\!=\!6$ Chern-Simons index at
large $N$. The comparison of $I^+$ for the $\mathcal{N}\!=\!6$ Chern-Simons theory and gravity
on $AdS_4\times S^7$ is studied in detail in \cite{Kim:2009wb}. So all analytic and
numerical results there could be used in our context to show the agreement of the index
of this section. In the next subsection, we shall prove that the the agreement of the
gauge-gravity large $N$ indices for the $\mathcal{N}\!=\!6$ theory directly implies the
agreement of the large $N$ index of our system for $k\!=\!1$..

From the absence of these fundamental degrees of freedom, the integrand in $I^\pm$ is
invariant under the common shift of the holonomies $\alpha,\tilde\alpha$, implying that
it acquires contribution only from monopoles satisfying
\begin{equation}
  \sum n_i=\sum\tilde{n}_i\ .
\end{equation}
This decoupling of the diagonal $U(1)$ in $U(N)\times U(N)$ was an exact property
in the $\mathcal{N}\!=\!6$ Chern-Simons-matter theory, while here it is true only
in the large $N$ limit.

\subsection{Gravity}

The complete Kaluza-Klein spectrum on $AdS_4\times N^{010}$ was obtained in \cite{Fre':1999xp}.
To calculate the index, it suffices to consider the fields in the short multiplets.\footnote{
By `fields' we mean fields in $AdS_4$. This is obtained by decomposing $OSp(3|4)$
representations into the representations of the conformal group, where each conformal
representation comes from a 4 dimensional field.}
There exists a tower of short or massless graviton multiplets of $OSp(3|4)\times SU(3)$
with $M_1\!=\!M_2\!=\!p$, $J_0\!=\!p$ and $p\!\geq\!0$, where the lowest multiplet with $p\!=\!0$
is the massless multiplet. $M_1,M_2$ denote the two parameters of the $SU(3)$
representation as used in \cite{Fre':1999xp}, and $J_0$ denotes the $SU(2)_R$ Casimir of the
primary. There also exists a tower of short gravitino multiplets, with $M_1\!=\!p$,
$M_2\!=\!p\!+\!3$, $J_0\!=\!p\!+\!1$ and $p\geq 0$, all being massive. As these multiplets are
in complex representations of $SU(3)$ with $M_1\!\neq\!M_2$, the corresponding 4 dimensional
field is complex. Therefore, when calculating the index below, we should include the
contribution from the conjugate modes with $M_1\!=\!p\!+\!3$, $M_2\!=\!p$.
Also, a tower of short or massless vector multiplets comes with $M_1\!=\!M_2\!=\!p$,
$J_0\!=\!p$ and $p\geq 1$, where the multiplet with $p\!=\!1$ is massless and corresponds
to the gauge fields for the $SU(3)$ isometry. Finally, there appears another massless
vector multiplet with $M_1\!=\!M_2\!=\!0$, $J_0\!=\!1$ which corresponds to the baryonic
$U(1)$ symmetry, under which the M5-brane wrapped on a 5-cycle in $N^{010}$ is charged.
From these multiplets, the fields (or representations of the conformal group) which satisfy
the BPS relation $\epsilon\!=\!h_0\!+\!j_3$ are listed in Table \ref{BPS-field}.
\begin{table}[t!]
$$
\begin{array}{c|ccccc}
  \hline&j_3&\epsilon&h_0&\epsilon\!+\!j_3&(M_1,M_2)\\
  \hline p\geq 0&\frac{3}{2}&p+\frac{5}{2}&p+1&p+4&(p,p)\\
  p\geq 0&1&p+2&p+1&p+3&(p,p)\\
  \hline p\geq 0&1&p+3&p+2&p+4&(p,p+3),(p+3,p)\\
  p\geq 0&\frac{1}{2}&p+\frac{5}{2}&p+2&p+3&(p,p+3),(p+3,p)\\
  \hline p\geq 1&\frac{1}{2}&p+\frac{1}{2}&p&p+1&(p,p)\\
  p\geq 1&0&p&p&p&(p,p)\\
  \cdot&\frac{1}{2}&\frac{3}{2}&1&2&(0,0)\\
  \cdot&0&1&1&1&(0,0)\\
  \hline
\end{array}
$$
\caption{Fields saturating the BPS bound for $N^{010}$}\label{BPS-field}
\end{table}

Identifying $SU(3)$ Cartans with the $SU(2)_F$ Cartan and $U(1)_B$
from theory theory as \cite{Gaiotto:2009tk}
\begin{equation}
  h_1={\rm diag}\left(\frac{1}{2},-\frac{1}{2},0\right)\ ,\ \
  h_2={\rm diag}\left(\frac{1}{6},\frac{1}{6},-\frac{1}{3}\right)\ ,
\end{equation}
the single particle index from gravity is given by
\begin{align}
&I_{\rm sp}(x,y_1 ,y_2)=
\frac{1}{1-x^2}\big{[}\sum_{l=0}^\infty (x^{l+3}-x^{l+4})\chi^{(l ,l)}_{SU(3)}(y_1 ,y_2) \nonumber
\\
&+\sum_{l=0}^\infty (x^{l+4}-x^{l+3})\big{(}\chi^{(l ,l+3)}_{SU(3)}(y_1 ,y_2)+\chi^{(l+3 ,l)}_{SU(3)}(y_1 ,y_2)\big{)} +\sum_{l=1}^\infty (x^{l}-x^{l+1})\chi^{(l ,l)}_{SU(3)}(y_1 ,y_2) + (x-x^2)  \big{]}, \nonumber
\\
&= \frac{1}{1+x}\big{[}x+ \sum_{l=0}^\infty x^{l+3} \big{(} \chi^{(l ,l)}_{SU(3)}(y_1 ,y_2)- \chi^{(l ,l+3)}_{SU(3)}(y_1 ,y_2)-\chi^{(l+3 ,l)}_{SU(3)}(y_1 ,y_2)\big{)} +\sum_{l=1}^\infty x^l \chi^{(l ,l)}_{SU(3)}(y_1 ,y_2)  \big{]}. \label{single graviton index}
\end{align}
The $SU(3)$ character $\chi_{SU(3)}^{M_1,M_2}(y_1,y_2)$ for an irreducible
representation is given by
\begin{align}
&\chi^{(M_1 ,M_2)}_{SU(3)}(y_1 ,y_2)\equiv{\rm tr}_{(M_1,M_2)}y_1^{h_1}y_2^{h_2}=
\frac{\rm (numerator)}{\rm (denominator)}, \nonumber
\\
&({\rm numerator})=y_1^{-\frac{1}2(M_1 +M_2)} y_2^{-\frac{1}6(2M_1 +M_2)}\big{[}y_1^{\frac{1}2(M_1+1)}-y_1^{\frac{1}2(M_1+3)+M_2} -y_2^{\frac{1}2(M_1+1)} \nonumber
\\
&+y_1^{(M_1+M_2+2)} y_2^{\frac{1}2(M_1+1)} +y_1^{\frac{1}2(M_2+1)} y_2^{\frac{1}2(M_1+M_2+2)}-y_1^{\frac{1}2(2M_1+M_2+3)} y_2^{\frac{1}2(M_1+M_2+2)}\big{]}, \nonumber
\\
&({\rm denominator})=(y_1 -1)\big{[}y_2^{\frac{1}2}+y_1 y_2^{\frac{1}{2}}- y_1^{\frac{1}2} (y_2+1)\big{]}. \label{character of SU(3)}
\end{align}
From this single particle index, the full index over the gravity states is given by
\begin{equation}
  I(x,y_1 ,y_2) =\exp\left[\sum_{n=1}^\infty \frac{1}{n}I_{\rm sp}(x^n , y_1^n ,y_2^n)\right]\ .
\end{equation}
Below, we compare the field theory and gravity indices when $k\!=\!1$. The case with
$k\!\neq\!1$ is studied in the next subsection, with D6-brane contributions taken into account.

For the field theory side at $k\!=\!1$, one either sets $(m_1,m_2)=(1,0)$ or $(m_1,m_2)=(0,1)$
to study the field theory dual of $N^{010}$ (rather than other tri-Sasakian spaces).
In the `D6-brane' picture (although one needs the full M-theory at $k\!=\!1$), these two cases
correspond to having different $\mathbb{Z}_2$ valued Wilson lines on the worldvolume of D6-brane.
Geometrically uplifting D6-branes to $N^{010}$, this should lift to some kind of $\mathbb{Z}_2$
valued holonomy of the 3-form potential of M-theory. As such holonomy does not affect
the spectrum of gravity fields, the two field theories would give identical large $N$
spectrum that we consider in this section. This was what we encountered for the field theory
in the previous subsection. We thus consider the case $(m_1,m_2)=(1,0)$ for definiteness.

As the field theory index is factorized as $I^{(0)}I^+I^-$ where the three factors are neutral
or positively/negatively charged in $h_2$ $U(1)_B$ charge, we can make the same decomposition of
$I_{\rm sp}(x,y_1,y_2)$ on the gravity side and compare the three factors separately. The
single particle index in the neutral sector is given by
\begin{align}
&I^{(0)}_{\rm sp}(x,y_1)=\frac{1}{2\pi i }\oint\frac{d y_2}{y_2}I_{\rm sp}(x,y_1 ,y_2^2)\nonumber\\
&= \frac{x }{y_1-x}+\frac{x y_1}{1-x y_1}+\frac{2 x}{1-x} -\frac{2
x^2}{1-x^2}+ \frac{x y_1}{\left(y_1-x\right) \left(1-x
y_1\right)}. \label{gravity I(0)}
\end{align}
The sum of first four terms is the same as the single particle index on $AdS_4\times\mathbb{CP}^3$.
The last term may be interpreted as a contribution from single `D6-brane' wrapping
$\mathbb{RP}^3\subset\mathbb{CP}^3$, as explained in more detail in the next subsection.
The neutral sector `single particle index' $I^{(0)}_{single}$ of (\ref{field I(0)})
from field theory completely agrees with $I^{(0)}_{\rm sp}$ of (\ref{gravity I(0)}) as one
inserts $m_1\!=\!1$, $m_2\!=\!0$. This shows the agreement in the neutral sector.

To study the sector with positive $h_2$, one generally has to rely on numerical analysis like
\cite{Kim:2009wb}, as we do not know how to calculate holonomy integral (\ref{formula for I'})
analytically. As we mentioned in the previous subsection, the field theory $I^+$ is almost
identical for our $\mathcal{N}\!=\!3$ theory and the $\mathcal{N}\!=\!6$ theory so that all
results (analytic or numerical) known for the latter case can be borrowed to study the former.

For instance, in the simplest sector with minimal positive value $h_2\!=\!\frac{1}{2}$
which was treated analytically in \cite{Kim:2009wb}, the corresponding gravity index in
this sector comes from one particle states with $h_2\!=\!\frac{1}{2}$,
\begin{equation}\label{gravity I(1/2)}
 I^{(1/2)}_{\rm sp}(x,y_1)=\oint\frac{d\sqrt{y_2}}{2\pi i\sqrt{y_2}}y_2^{-1/2}I_{\rm sp}(x,y_1,y_2) =-\frac{x \left(1+y_1\right) \left(-x^2+y_1-x^2 y_1+2 x^3 y_1-x^2 y_1^2\right)}{(-1+x) (1+x) \left(x-y_1\right) \sqrt{y_1} \left(-1+x y_1\right)}\ ,
\end{equation}
which is the $\mathcal{O}(y_2^{1/2})$ contribution to $I^+(x,y_1,y_2)$.
The corresponding field theory index comes with nonzero fluxes $n_1\!=\!\tilde{n}_1\!=\!1$ and
all other fluxes being zero. Here one obtains from (\ref{formula for I'})
\begin{align}
 I^+_{(1)(1)}&=
  x^{k/2}\int_0^{2\pi} \frac{d\alpha d\tilde{\alpha}}{(2\pi)^2}  e^{ik (\alpha - \tilde{\alpha})}\exp \big{[}\sum_{n=1}^\infty \frac{1}n \big{(} (1-x^{2n}) \big {(} f^+(\cdot^n) e^{in (\tilde{\alpha} - \alpha)} +f^-(\cdot^n) e^{in (\alpha - \tilde{\alpha})}\big{)}  + 2x^{2n} \big{)} \big{]} \nonumber \\
 & = x^{k/2}\oint \frac{dz}{2 \pi i z} z^{-k}
 \frac{\left(1-\frac{x^{3/2}}{z \sqrt{y_1}}\right) \left(1-\frac{x^{3/2} z}{\sqrt{y_1}}\right) \left(1-\frac{x^{3/2} \sqrt{y_1}}{z}\right) \left(1-x^{3/2} z \sqrt{y_1}\right)}
 {\left(1-x^2\right)^2 \left(1-\frac{\sqrt{x}}{z \sqrt{y_1}}\right) \left(1-\frac{\sqrt{x} z}{\sqrt{y_1}}\right) \left(1-\frac{\sqrt{x} \sqrt{y_1}}{z}\right) \left(1-\sqrt{x} z \sqrt{y_1}\right)}.
\end{align}
As emphasized, this is the same as the large $N$ $\mathcal{N}\!=\!6$ index apart from the extra
factor $x^{k/2}$. At $k\!=\!1$, the contour integral can be performed to yield
\begin{equation}\label{field I(1/2)}
 I^+_{(1)(1)}\left.\frac{}{}\right|_{k\!=\!1}=-\frac{x \left(1+y_1\right)\left(-x^2+y_1-x^2 y_1+2 x^3 y_1-x^2 y_1^2\right)}{(-1+x) (1+x) \left(x-y_1\right) \sqrt{y_1} \left(-1+x y_1\right)}\ ,
\end{equation}
in perfect agreement with (\ref{gravity I(1/2)}).

Pushing this sort of analysis further, we assume the agreement between the gauge/gravity
$\mathcal{N}\!=\!6$ indices at large $N$ (shown in \cite{Kim:2009wb} up to three monopoles in
various sectors) and show that this directly implies the agreement of the large $N$ indices here.
Firstly, in the positive flux sector, the $\mathcal{N}\!=\!6$ index $I^+_{ABJM}(x,y_1,y_2,y_3)$
is related to our $I^+$ by
\begin{equation}\label{relation}
  I^+(x,y_1,y_2)=I^+_{ABJM}(x,y_1,y_2\!=\!1,y_3\!=\!y_2x)\ .
\end{equation}
$y_1,y_2$ are chemical potentials of $SU(2)\times SU(2)\subset SO(6)$ global symmetry of the
latter system in the notation of \cite{Kim:2009wb}. As only the former $SU(2)$ is the symmetry
in our case, we set $y_2\!=\!1$. Also,
as the charge conjugate to $y_3$ in the latter case is just given by $\frac{1}{2}\sum_in_i$
from positive fluxes, the replacement $y_3\rightarrow y_2x$ yields the desired extra factor
$x^{\frac{1}{2}\sum n_i}$ of zero point energy. Now the agreement between
$\mathcal{N}\!=\!6$ large $N$ gauge/gravity indices (which we accept) means
$I^+_{ABJM}=\exp\left[\sum_{n=1}^\infty\frac{1}{n}I^+_{\rm sp}(\cdot^n)_{S^7}\right]$,
while we would like to show
\begin{equation}
  I^+=\exp\left[\sum_{n=1}^\infty\frac{1}{n}I^+_{\rm sp}(\cdot^n)_{N^{010}}\right]\ .
\end{equation}
Here, $(I^+_{\rm sp})_{S^7}$ and $(I^+_{\rm sp})_{N^{010}}$ denote single particle gravity
indices in appropriate backgrounds. Thus, from (\ref{relation}), it suffices for us to show
\begin{equation}\label{gravity-relation}
  I^+_{\rm sp}(x,y_1,y_2)_{N^{010}}=
  I^+_{\rm sp}(x,y_1,y_2\!\rightarrow\!1,y_3\!\rightarrow\!y_2x)_{S^7}.
\end{equation}
Using computer, one can explicitly show to all order that
\begin{eqnarray}
I^{+}_{\rm sp} (x,y_1,y_2)_{N^{010}}&=&
\sum_{m=1}^{\infty} I^{(m/2)}_{\rm sp} (x, y_1 )_{N^{010}}y_2^{m/2}\nonumber\\
&=&\sum_{m=1}^{\infty}\left([Res_{y_2 \rightarrow  x \sqrt{y_1}}  +Res_{y_2 \rightarrow  x/\sqrt{y_1}}] \frac{I_{\rm sp}(x,y_1 ,y_2^2)_{N^{010}}y_2^{m}}{y_2}\right) y_2^{m/2}, \nonumber
\\
&=&\sum_{m=1}^\infty \big{(}\frac{(x^2 y_1 y_2 )^{m/2} (x - y_1 - x^3 y_1 + x^2 y_1^2)}{(1-y_1)(1-x^2)(1- x y_1)} +(y_1 \rightarrow y_1^{-1}) \big{)}.
\end{eqnarray}
and
\begin{eqnarray}
I^{(+)}_{\rm sp} (x, y_1,1,y_3)_{S^7}&=&
\sum_{m=1}^{\infty} I^{(m/2)}_{\rm sp}(x, y_1 , 1)_{S^7}y_3^{m/2}\nonumber
\\
&=&\sum_{m=1}^{\infty}\left([Res_{y_3 \rightarrow \sqrt{x y_1}}  +Res_{y_2 \rightarrow \sqrt{x /y_1}}] \frac{I_{\rm sp}(x,y_1 ,1, y_3^2)_{S^7} y_3^{m}}{y_3}\right) y_3^{m/2}, \nonumber
\\
&=&\sum_{m=1}^\infty \big{(}\frac{(x y_1 y_3)^{m/2} (x - y_1 - x^3 y_1 + x^2 y_1^2)}{(1- y_1 )(1-x^2)(1-x y_1)} +(y_1 \rightarrow y_1^{-1}) \big{)}.
\end{eqnarray}
The apparent $y_1\rightarrow 1$ singularity in each term cancels with a pair term with
$y_1\rightarrow y_1^{-1}$ replacement. From these, (\ref{gravity-relation}) is obvious,
which proves the agreement between the gauge/gravity indices for $N^{010}$.

\subsection{D6-branes}

When $k\!\neq\!1$, $N^{010}/\mathbb{Z}_k$ is singular so that there appear
contributions beyond the supergravity approximation, coming from light degrees of
freedom localized on the fixed points of $\mathbb{Z}_k$. Similar analysis was done
for $\mathcal{N}\!=\!4$ M2-brane models in \cite{Imamura:2009hc,Imamura:2010sa}.

In our case, one can use the type IIA D6-brane picture of the field theory model
when $k\!\gg\!1$.\footnote{This $D6$-brane picture turns out to be appropriate for studying
the index even for small $k$ (including $k\!=\!1$), presumably due to simplifications
coming from supersymmetry.} In this picture, one starts from the $\mathcal{N}\!=\!6$
Chern-Simions-matter theory for M2-branes and add fundamental flavors by adding D6-branes
wrapping $AdS_4\times\mathbb{RP}^3\subset AdS_4\times\mathbb{CP}^3$
\cite{Hohenegger:2009as,Gaiotto:2009tk,Hikida:2009tp}. The type IIA index from gravity can
then be computed by adding the single particle index from the bulk modes in
$AdS_4\times\mathbb{CP}^3$ and the modes on the D6-brane worldvolumes. As
$\pi_1(\mathbb{RP}^3)=\mathbb{Z}_2$, there appears two types of D6-branes with different
values of discrete Wilson line on the worldvolume \cite{Hikida:2009tp}. As mentioned in the
previous subsection, the single particle index on $AdS_4\times\mathbb{CP}^3$ is given by \cite{Bhattacharya:2008bja}
\begin{equation}
  I_{\rm sp}^{\mathbb{CP}^3}(x,y_1)=\frac{x}{y_1-x}+\frac{x y_1}{1-x y_1}
  +\frac{2 x}{1-x}-\frac{2x^2}{1-x^2}
\end{equation}
where $y_1$ is the chemical potential for the Cartan of $SU(2)_F\subset SU(4)$ unbroken
by the D6-branes.

The open string degrees of freedom on D6-branes are described by a
7 dimensional supersymmetric Yang-Mills theory on $AdS_4\times\mathbb{RP}^3$. The quadratic
part of this action on $AdS_4\times S^3$ was obtained in \cite{Imamura:2010sa}, which in our
case can be derived from the DBI action with Wess-Zumino term, as studied in \cite{Hikida:2009tp}.
Let $m_1$ and $m_2$ be the number of D6-branes which support two possible $\mathbb{Z}_2$ valued
Wilson lines. These numbers are identified with the number of two fundamental hypermultiplets
in the field theory. The fields in a vector multiplet are: 7 dimensional gauge field $A_\mu$,
three real scalars $\phi_i$ ($i\!=\!1,2,3$), gauginos $\lambda$. The symmetry of this
worldvolume theory is: $U(m_1)\times U(m_2)$ gauge symmetry, $SO(4)\equiv SU(2)_1\times
SU(2)_2$ isometry on $S^3$ or $\mathbb{RP}^3$, $SO(3)\sim SU(2)_3$ symmetry transverse to
the D6-branes.\footnote{In \cite{Imamura:2010sa}, the $SU(2)$ symmetries are named
$SU(2)_{1,2,3}=SU(2)_R^\prime,SU(2)_F^\prime,SU(2)_R$, respectively.} Viewing the four
bi-fundamental scalars $A_1,B_1^\dag$ ,$A_2,B_2^\dag$ as spanning the $\mathbb{C}^4$ space
(at least for $N\!=\!1$), the embedding condition for $\mathbb{RP}^3$ in $\mathbb{CP}^3$
should be specifying $\mathbb{C}^2=\mathbb{R}^4\subset\mathbb{C}^4$ embedding by
taking one combination of the two hypermultiplets to zero, as the D6-brane embedding should
be compatible with $SU(2)_R$ R-symmetry. Therefore, one of the two hypermultiplets
span transverse directions, acted by $SU(2)_3$, while another one is acted by
$SU(2)_1\times SU(2)_2$. $SU(2)_{1,2,3}$ are symmetries of the low energy theory on D6-branes,
and some of them are broken in the whole theory. In particular, the diagonal combination
of $SU(1)_1$ and $SU(2)_3$ is to be identified with our $SU(2)_R$ symmetry, while $SU(2)_2$
is our $SU(2)_F$ symmetry.

The Kaluza-Klein spectrum of the 7 dimensional modes on $\mathbb{RP}^3$ or $S^3$ are
summarized in Table \ref{D6-KK}, following \cite{Imamura:2010sa}.
\begin{table}[t!]
$$
\begin{array}{c|ccccc}
  \hline{\rm fields}&SU(2)_j&SU(2)_3&SU(2)_1&SU(2)_2&\epsilon\\
  \hline \phi_i&0&1&s&s&s+2\\ \hline A_\mu&0&0&s+1&s&s+1\\&1&0&s&s&s+2\\&0&0&s-1&s&s+3\\
  \hline \lambda&\frac{1}{2}&\frac{1}{2}&s+\frac{1}{2}&s&s+\frac{3}{2}\\
  &\frac{1}{2}&\frac{1}{2}&s-\frac{1}{2}&s&s+\frac{3}{2}\\ \hline
\end{array}
$$
\caption{Spectrum of 7 dimensional vector supermultiplets on
$AdS_4\times S^3$ or $\mathbb{RP}^3$}\label{D6-KK}
\end{table}
For $S^3$, the number $s$ is half an integer. For $\mathbb{RP}^3$, $s$ is an integer
by the $\mathbb{Z}_2$ projection when the open strings connect D6-branes with same
Wilson line. There are $m_1^2+m_2^2$ such vector multiplets. On the other hand,
$s$ is half an odd integer when the open strings connect D6-branes of different Wilson
lines, as there are half-integral shifts of the spectrum due to their coupling to
nonzero Wilson lines \cite{Hikida:2009tp}. The modes saturating the BPS bound
come from the second and fifth lines of Table \ref{D6-KK}. The single particle index
(or to be more precise, the mode index) from the open strings is
\begin{eqnarray}
  I_{\rm sp}^{D6}&=&\frac{m_1^2+m_2^2}{1-x^2}\sum_{s=0}^\infty
  \chi_s^{SU(2)_F}(y_1)\left[x^{s+1}-x^{s+2}\right]+\frac{2m_1m_2}{1-x^2}
  \sum_{s=\frac{1}{2}}^\infty\chi_s^{SU(2)_F}(y_1)\left[x^{s+1}-x^{s+2}\right]\nonumber\\
  &=&\frac{(m_1^2+m_2^2)x}{(1-xy_1)(1-xy_1^{-1})}+\frac{2m_1m_2x^{3/2}(y_1^{1/2}+
  y_1^{-1/2})}{(1-xy_1)(1-xy_1^{-1})}\ ,
\end{eqnarray}
where $\chi_s^{SU(2)_F}(y_1)=\frac{y_1^{(s+1)/2}-y_1^{-(s+1)/2}}{y_1^{1/2}-y_1^{-1/2}}$.
From this, one finds that $I_{\rm sp}^{\mathbb{CP}^3}+I_{\rm sp}^{D6}$ perfectly agrees with
$I_{single}^{(0)}$ in (\ref{field I(0)}).
In particular, extending the above result to $k\!=\!1$, we confirm that
the $U(1)_B$ neutral part (\ref{gravity I(0)}) of the index on $N^{010}$ is the
gravity index on $AdS_4\times\mathbb{CP}^3$ with one D6-brane.

\section{The index for M2-branes probing $C(Q^{111})$}

\subsection{Gravity}

The gravity spectrum on $AdS_4\times Q^{111}$ has been analyzed in \cite{Merlatti:2000ed}.
The index will acquire contribution only from the short multiplets. In \cite{Merlatti:2000ed},
$15$ infinite towers of short multiplets ($1$ tower of short graviton, $9$ towers of short gravitino,
$4$ towers of short vector multiplets and $1$ tower of hypermultiplet) plus $6$ massless multiplets
($1$ massless graviton, $5$ massless vectors) belong to this category.

As far as we are aware of, only the massless multiplets as well as
some low-lying entries in the hypermultiplet tower (containing
states dual to chiral ring operators) have been studied or
cross-checked in the literature. From the anlaysis of the field
theory dual of \cite{Jafferis:2009th} proposed for the $Q^{111}$, it
still turns out that the large $N$ field theory index agrees with
the gravity multiplets after discarding $6$ of the $9$ proposed
towers of gravitino multiplets in \cite{Merlatti:2000ed}, and $1$ of
the $4$ proposed towers of short vector multiplets with $SU(2)^3$
charge $(k/2,k/2,k/2)$. It could be that one might have to carefully
re-examine the gravity spectrum of \cite{Merlatti:2000ed}, which is
beyond the scope of this work.

In Table \ref{Q111}, we present the set of gravity multiplets which will be compared to
the field theory results in the next subsection.
\begin{table}[t!]
$$
\begin{array}{c|cc|ccc}
  \hline &\epsilon+j_3&j_3&SU(2)_1&SU(2)_2&SU(2)_3\\
  \hline k\geq 1&k+4&3/2&k/2&k/2&k/2\\
  \hline k\geq 0&k+4&1&k/2&k/2+1&k/2+1\\
  &k+4&1&k/2+1&k/2&k/2+1\\ &k+4&1&k/2+1&k/2+1&k/2\\
  \hline k\geq 1&k+2&1/2&k/2+1&k/2&k/2\\
  &k+2&1/2&k/2&k/2+1&k/2\\ &k+2&1/2&k/2&k/2&k/2+1\\
  \hline k\geq 1&k&0&k/2&k/2&k/2\\
  \hline \cdot &4&3/2&0&0&0\\
  \hline \cdot &2&1/2&1&0&0\\
  \cdot&2&1/2&0&1&0\\ \cdot&2&1/2&0&0&1\\ \hline
\end{array}
$$
\caption{Gravity spectrum, ignoring $6$ towers of short gravitino
and $1$ tower of short vector multiplet of \cite{Merlatti:2000ed}.
The first line from short graviton, second-fourth lines from short gravitino,
fifth-seventh lines from the short vector, eighth line from hypermultiplets.
The last four lines are from massless graviton and vector multiplets.}\label{Q111}
\end{table}

The single particle index over the above fields is
\begin{eqnarray}
  \hspace*{-1.7cm}(1-x^2)I_{\rm sp}(x,y,y_2)&=&(1-x^2){\rm tr}\left[(-1)^Fx^{\epsilon+j_3}y^{2J_1}y_2^{J_2+J_3}
  \right]=-\sum_{k=1}^\infty\chi_{k/2}(y)\chi_{k/2}(y_2^{1/2})^2x^{k+4}-x^4\\
  \hspace*{-1cm}&&\hspace{-2cm}+\sum_{k=0}^\infty\left(\chi_{k/2+1}(y_2^{1/2})^2
  \chi_{k/2}(y)+2\chi_{k/2}(y_2^{1/2})\chi_{k/2+1}(y_2^{1/2})\chi_{k/2+1}(y)\right)x^{k+4}\nonumber\\
  \hspace*{-1cm}&&\hspace{-2cm}-\sum_{k=1}^\infty\left(\chi_{k/2}(y_2)^2\chi_{k/2+1}(y)+2\chi_{k/2}(y_2^{1/2})
  \chi_{k/2+1}(y_2^{1/2})\chi_{k/2}(y)\right)x^{k+2}
  -(\chi_1(y)+2\chi_1(y_2^{1/2})+2)x^2\nonumber\\
  I_{\rm sp}&=&
    \frac{x\left(y_2^{1/2}+y_2^{-1/2}\right)^2\left((y+y^{-1})-x(y_2^{1/2}+y_2^{-1/2})^2
    +x^2(y+y^{-1})\right)}{\left(1-xy_1y_2\right)\left(1-xy/y_2\right)\left(1-xy_2/y\right)
    \left(1-x/(yy_2)\right)}\nonumber
\end{eqnarray}
with $\chi_j(y)=\frac{y^{2j+1}-y^{-2j-1}}{y-y^{-1}}$,
where $J_{1,2,3}$ is the Cartan of $SU(2)_{1,2,3}$, respectively, and the combination
$J_2+J_3$ is to be identified with the $U(1)_B$ charge carried by the monopoles
$\sum_i n_i=\sum_i\tilde{n}_i$ in the large $N$ limit. $y$ would later be identified with
$y_1$ in field theory as $y=y_1^{1/2}$.

Expanding this result in powers of $y_2$, one obtains the single paraticle indices with
definite $U(1)_B$ (or $J_2+J_3$) charges as follows:
\begin{eqnarray}\label{gravity-single-q111}
  I^{(0)}_{\rm sp}&=&2\left(\frac{xy_1^{1/2}}{1-xy_1^{1/2}}+\frac{xy_1^{-1/2}}{1-xy_1^{-1/2}}
  -\frac{x^2}{1-x^2}\right)\\
  I^{(1)}_{\rm sp}&=&\frac{x \left(y+y^{-1} + 4 x^3 + x (-y^{-1} + y)^2 - 3 x^2 (y^{-1} + y)\right)}
  {(1-x^2) (1-xy^{-1})(1-x y)}\nonumber\\
  I^{(2)}_{\rm sp}&=&\frac{x^2 (x^4 y^3 - 3 x^3 (y^2 + y^4) - y (1 + y^2 + y^4) +
   3 x^2 (y + y^5) - x (1 - 2 y^2 - 2 y^4 + y^6))}{(-1 + x^2) (x -
   y) y^2 (-1 + x y)}\ ,\nonumber
\end{eqnarray}
and so on ($y=y_1^{1/2}$).

\subsection{Field theory}

The field theory model proposed for M2-branes on $C(Q^{111})$ adds four fundamental chiral
supermultiplets to the $\mathcal{N}\!=\!6$ theory: two of them $q_1,q_2$ are anti-fundamental
in the first gauge group, while the other two $\tilde{q}_1$, $\tilde{q}_2$ are fundamental in
the second gauge group. This model
is proposed to described M2-branes on $C(Q^{111})$ when the bare Chern-Simons level $k$ is $0$.
One should also add the following superpotential
\begin{equation}
  \Delta W =q_1A_1\tilde{q}_1+q_2A_2\tilde{q}_2
\end{equation}
to the $\mathcal{N}\!=\!6$ theory.

In the $U(1)\times U(1)$ theory, the OPE of the monopole operators $T$, $\tilde{T}$ with charges
$(1)(1)$ and $(-1)(-1)$ is given by $T\tilde{T}=A_1A_2$, which can be solved as
\begin{equation}
  A_1=a_1b_2\ ,\ \ A_2=b_1a_2\ ,\ \ T=a_1b_1\ ,\ \ \tilde{T}=a_2b_2\ .
\end{equation}
This solution is invariant under $(a_1,a_2,b_1,b_2)\rightarrow(\lambda a_1,\lambda a_2,
\lambda^{-1}b_1,\lambda^{-1} b_2)$.
On the space of $(B_1,B_2,a_1,a_2,b_1,b_2)$, $\mathbb{C}^6//U(1)^2$ is given by
$(\mu^{-2}B_1,\mu^{-2}B_2,\mu\lambda a_1,\mu\lambda a_2,\mu\lambda^{-1}b_1,\mu\lambda^{-1}b_2)$,
or equivalently $(\lambda B_1,\lambda B_2,\mu a_1,\mu a_2,\nu b_1, \nu b_2)$ with $\lambda\mu\nu=1$,
giving the cone over $Q^{111}$. The three $SU(2)$ symmetries are  given by rotating $B_1,B_2$ or
$a_1,a_2$ or $b_1,b_2$.

In the non-Abelian field theory, only the $SU(2)$ rotating $B_1,B_2$ is manifest.
For the remaining $SU(2)$'s only the Cartans could be manifest. Under the Cartans of second and
third $SU(2)$'s rotating $a_1,a_2$ and $b_1,b_2$, $A_{1,2}$ carry charge $\pm\frac{1}{2}$
and $\mp\frac{1}{2}$. Also, the monopole charge $\frac{1}{2}\sum n_i$($=\frac{1}{2}\sum\tilde{n}_i$
in the large $N$ limit) would contribute to this charge. Phase rotations of $A_{1,2}$ explained
above for are not symmetries unless accompanied by appropriate rotations of fundamental fields,
to make the superpotential invariant. For simplicity, we only weight the states with
the sum of two $SU(2)$ Cartans in this paper, carried by the monopole charges.

The bosonic fields and charges are listed in Table \ref{Q111-field}. Note that
the R-charges of fields are constrained to have the superpotential marginal, after
which only one parameter $b$ is left. This parameter is expected to be $b\!=\!\frac{1}{3}$,
for instance from the study of baryonic states from the gravity dual. This can be determined
by studying the extremization of the partition function \cite{Jafferis2}, say in the
large $N$ limit \cite{Herzog:2010hf}, or by comparing the $\mathcal{O}(N)$ energy spectrum
of the index with gravity. In this paper, we only study the low energy spectrum in the large
$N$ limit and work with the parameter $b$ unfixed. It will turn out that the low energy
large $N$ index does not depend on $b$, after imposing the gauge invariance constraint (holonomy
integrals, to be explained shortly).
\begin{table}[t!]
$$
\begin{array}{c|ccc|c}
  \hline {\rm fields}&SU(2)_1&SU(2)_2+SU(2)_3&R&U(N)\times U(N)\\
  \hline B_1&+1/2&0&b=1/3&(\bar{N},N)\\
  B_2&-1/2&0&b=1/3\\ \hline A_1&0&0&a=1-b&(N,\bar{N})\\
  A_2&0&0&a=1-b\\
  \hline q_1&0&0&\frac{1+b}{2}&(\bar{N},1)\\
  q_2&0&0&\frac{1+b}{2}\\ \hline\tilde{q}_1&0&0&\frac{1+b}{2}&(1,N)\\
  \tilde{q}_2&0&0&\frac{1+b}{2}&\\ \hline T=(1)(1)&0&1&1-b\\
  \tilde{T}=(-1)(-1)&0&-1&1-b\\ \hline
\end{array}
$$
\caption{Fields and charges: $SU(2)_{1,2,3}$ denote the Cartan charges}\label{Q111-field}
\end{table}

From the general expressions in \cite{Imamura:2011su}, one finds the following
index with monopoles charges $\{n_i\}$, $\{\tilde{n}_i\}$:
\begin{eqnarray}
  I&=&\frac{x^{\epsilon_0}}{({\rm symmetry})}\int\left[\frac{d\alpha d\tilde\alpha}
  {(2\pi)^2}\right]e^{ib_0}\exp\left[\sum_{i,j=1}^N\sum_{n=1}^\infty\frac{1}{n}
  x^{n|n_i-\tilde{n}_j|}\left(f^+(\cdot^n)e^{-in(\alpha_i-\tilde\alpha_j)}+f^-(\cdot^n)
  e^{in(\alpha_i-\tilde\alpha_j)}\right)\right]\nonumber\\
  &&\exp\left[\sum_{i=1}^N\sum_{n=1}^\infty\frac{1}{n}
  x^{n|n_i|}\left(f_B(\cdot^n)e^{-in\alpha_i}+f_F(\cdot^n)e^{in\alpha_i}\right)+
  (n_i,\alpha_i\rightarrow\tilde{n}_i,-\tilde\alpha_i)\right]\nonumber\\
  &&\exp\left[-\sum_{i\neq j}^N\sum_{n=1}^\infty\frac{1}{n}\left(x^{n|n_i-n_j|}e^{-in(\alpha_i-\alpha_j)}+
  (n_i,\alpha_i\rightarrow\tilde{n}_i,\tilde\alpha_i)\right)\right]
\end{eqnarray}
\begin{eqnarray}
  \epsilon_0&=&\sum_{i,j=1}^N|n_i-\tilde{n}_j|-\sum_{i<j}\left(|n_i-n_j|+
  |\tilde{n}_i-\tilde{n}_j|\right)+\frac{1-b}{2}\sum_{i=1}^N\left(|n_i|+|\tilde{n}_i|
  \right)\nonumber\\
  b_0&=&\sum_{i=1}^N\left(|n_i|\alpha_i-|\tilde{n}_i|\tilde\alpha_i\right)\ .
\end{eqnarray}
\begin{eqnarray}
  f^+&=&\frac{x^b}{1-x^2}\left(y_1^{1/2}+y_1^{-1/2}\right)-\frac{2x^{1+b}}{1-x^2}
  \ \ \ ({\rm from}\ B_a^\dag,\psi_{A_a})\nonumber\\
  f^-&=&\frac{2x^{1-b}}{1-x^2}-\frac{x^{2-b}}{1-x^2}\left(y_1^{1/2}+y_1^{-1/2}\right)
  \ \ \ ({\rm from}\ A_a^\dag,\psi_{B_a})\nonumber\\
  f_B&=&\frac{2x^{\frac{1+b}{2}}}{1-x^2}\ ,\ \ f_F=-\frac{2x^{\frac{3-b}{2}}}{1-x^2}\ .
\end{eqnarray}
Again after going to the large $N$ limit with energy and monopoles fixed to $\mathcal{O}(1)$,
the index factorizes as
\begin{equation}
  I_{N\rightarrow\infty}=I_0I^+I^-
\end{equation}
where
\begin{eqnarray}\label{I0-Q111}
  I_0&=&\prod_{n=1}^\infty\frac{1}{1-f^+(\cdot^n)f^-(\cdot^n)}\exp\left[\sum_{n=1}^\infty
  \frac{1}{n}\frac{2f_Bf_F+f^-f_B^2+f^+f_F^2}{1-f^+f^-}(\cdot^n)\right]\nonumber\\
  &=&\prod_{n=1}^\infty\frac{(1-x^{2n})^2}{\left(1-x^ny_1^{n/2}\right)^2\left(1-x^ny_1^{-n/2}\right)^2}\ .
\end{eqnarray}
Note that the second factor involving fundamental letter indices is simply $1$. The part $I^+$
depending on monopoles with positive flux is given by
\begin{eqnarray}
  \hspace*{-2cm}I^+&=&\frac{x^{\epsilon_0}}{({\rm symmetry})}\int \left[
  \frac{d\alpha d\tilde\alpha}{(2\pi)^2}\right]e^{i\sum_i(n_i\alpha-\tilde{n}_i\tilde\alpha_i)}\\
  \hspace*{-1.7cm}&&\hspace{-1.5cm}\exp\left[\sum_{i,j=1}\sum_{n=1}^\infty\frac{1}{n}
  \left(x^{n|n_i-\tilde{n}_j|}\!-\!
  x^{n|n_i|+n|\tilde{n}_j|}\right)\left(f^+(\cdot^n)e^{-in(\alpha_i-\tilde\alpha_j)}+
  f^-(\cdot^n)e^{in(\alpha_i-\tilde\alpha_j)}\right)\right]\nonumber\\
  \hspace*{-1.7cm}&&\hspace{-1.5cm}\exp\left[-\sum_{i,j}\sum_{n=1}^\infty
  \frac{1}{n}\left((1-\delta_{ij})x^{n|n_i-n_j|}-x^{n|n_i|+n|n_j|}\right)e^{-in(\alpha_i-\alpha_j)}
  -\sum_{i,j}\sum_{n=1}^\infty\frac{1}{n}\left((1-\delta_{ij})x^{n|\tilde{n}_i-\tilde{n}_j|}
  -x^{n|\tilde{n}_i|+n|\tilde{n}_j|}\right)e^{-in(\tilde\alpha_i-\tilde\alpha_j)}\right]\nonumber
\end{eqnarray}
Again the fundamental degrees disappear in $I^+$, apart from their contribution to
$\epsilon_0$ and $b_0$. On the gravity side, the flux $\sum n_i=\sum\tilde{n}_i$ in the large
$N$ corresponds to the sum of the Cartans of the two $SU(2)$'s. The $U(1)_B$ neutral graviton
contribution in (\ref{gravity-single-q111}) becomes
\begin{equation}
  \exp\left[\sum_{n=1}^\infty\frac{1}{n}I^{(0)}_{\rm sp}(x^n,y_1^n)\right]\ ,\ \
  I^{(0)}_{\rm sp}(x,y_1)=\frac{2xy_1^{1/2}}{1-xy_1^{1/2}}+
  \frac{2xy_1^{-1/2}}{1-xy_1^{-1/2}}-\frac{2x^2}{1-x^2}\ ,
\end{equation}
in perfect agreement with $I_0$ in (\ref{I0-Q111}) from field theory.

Let us now turn to the study of $U(1)_B$ charged states. From the structure of the
`quantum' Chern-Simons phase factor $e^{i(|n_i|\alpha_i-|\tilde{n}_i|\tilde\alpha_i)}$,
one can easily confirm that $I^+(x,y_1)=I^-(x,y_1)$. We thus study $I^+$ only.
We again start from the flux $(1)(1)$, which should agree with the single graviton
particles having unit $U(1)_B$ charge.
\begin{equation}
  I^+_{(1)(1)}(x,y_1)=x^{1-b}\int\frac{d\alpha}{2\pi}e^{i\alpha}
  \frac{\left(1-x^{1+b}e^{-i\alpha}\right)^2\left(1-x^{2-b}y_1^{1/2}e^{i\alpha}\right)
  \left(1-x^{2-b}y_1^{-1/2}e^{i\alpha}\right)}
  {\left(1-x^by_1^{1/2}e^{-i\alpha}\right)\left(1-x^by_1^{-1/2}e^{-i\alpha}\right)
  \left(1-x^{1-b}e^{i\alpha}\right)^2(1-x^2)^2}\ .
\end{equation}
After performing the contour integral, one obtains
\begin{equation}
  I^+_{(1)(1)}=-\frac{xy_1^{-1/2}}{1-x^2}-\frac{xy_1^{1/2}}{1-x^2}+\frac{2xy_1^{-1/2}}{1-xy_1^{-1/2}}
  +\frac{2xy_1^{1/2}}{1-xy_1^{1/2}}\ .
\end{equation}
This completely agrees with the gravity index $I^{(1)}_{\rm sp}$.

One can continue checking the agreement by studying the sector with two fluxes. One should
consider the fluxes $(2)(2)$, $(1,1)(1,1)$, $(2)(1,1)$, $(1,1)(2)$. One finds
\begin{eqnarray}
\hspace*{-2cm}I_{(2)(2)}&=&x^2 (1 + 1/y^2 + y^2) + x^3 (2/y^3 + 2 y^3) +
 x^4 (2/y^4 - 2/y^2 - 2 y^2 + 2 y^4)\\
 &&+ x^5 (2/y^5 + 2/y + 2 y + 2 y^5) +
 x^6 (-4 + 2/y^6 + 1/y^4 - 2/y^2 - 2 y^2 + y^4 + 2 y^6)\nonumber\\
 &&+ x^7 (2/y^7 - 4/y^3 + 4/y + 4 y - 4 y^3 + 2 y^7) +
 x^8 (-3 + 2/y^8 + 1/y^4 + 2/y^2 + 2 y^2 + y^4 + 2 y^8)+\cdots\nonumber\\
 \hspace*{-2cm}I_{(11)(11)}&=&x^2 (1 + 1/y^2 + y^2) + x^3 (2/y^3 + 2/y + 2 y + 2 y^3) +
 x^4 (2 + 5/y^4 + 2/y^2 + 2 y^2 + 5 y^4)\nonumber\\
 &&+ x^5 (6/y^5 + 6 y^5) + x^6 (5 + 9/y^6 - 1/y^4 + 2/y^2 + 2 y^2 - y^4 + 9 y^6)\nonumber\\
 &&+x^7 (10/y^7\!+\!4/y^3\!+\!2/y\!+\!2 y\!+\!4 y^3\!+\!10 y^7) +
 x^8 (-3 + 13/y^8 - 1/y^4 - 3/y^2 - 3 y^2 - y^4 + 13 y^8)+\cdots\nonumber\\
 \hspace*{-2cm}I_{(2)(11)}&=&I_{(11)(2)}=x^6 -x^7 (2/y+ 2 y) + x^8 (5 + 1/y^2 + y^2)+\cdots\nonumber
\end{eqnarray}
where $y=y_1^{1/2}$.
On the gravity side, the single particle index with $U(1)_B$ charge $2$ and two particle
contributions from particles with $U(1)_B$ charge $1$ are
\begin{eqnarray}
  \hspace*{-1cm}I^{(2)}_{\rm sp}(x,y)&=&x^2 (1 + 1/y^2 + y^2) + x^3 (2/y^3 + 2 y^3) +
  x^4 (2/y^4 - 1/y^2 - y^2 + 2 y^4)\\
  &&+ x^5 (2/y^5 + 2 y^5) +
  x^6 (2/y^6 - 1/y^2 - y^2 + 2 y^6) + x^7 (2/y^7 + 2 y^7)\nonumber\\
  && +x^8 (2/y^8 - 1/y^2 - y^2 + 2 y^8)+\cdots\nonumber\\
  \hspace*{-1cm}\frac{I^{(1)}_{\rm sp}(x^2,y^2)+I^{(1)}(x,y)_{\rm sp}^2}{2}&=&
  x^2 (1 + 1/y^2 + y^2) + x^3 (2/y^3 + 2/y + 2 y + 2 y^3) +
  x^4 (2 + 5/y^4 + 1/y^2 + y^2 + 5 y^4)\nonumber\\
  &&+x^5 (6/y^5 + 2/y + 2 y + 6 y^5) +x^6 (3 + 9/y^6 + 1/y^2 + y^2 + 9 y^6)\nonumber\\
  &&+x^7 (10/y^7 + 2/y + 2 y + 10 y^7) +
 x^8 (4 + 13/y^8 + 2/y^2 + 2 y^2 + 13 y^8)+\cdots\ .\nonumber
\end{eqnarray}
One finds that
\begin{equation}
  I_{(2)(2)}+I_{(11)(11)}+ I_{(2)(11)}+I_{(11)(2)}=I^{(2)}_{\rm sp}(x,y)
  +\frac{I^{(1)}_{\rm sp}(x^2,y^2)+I^{(1)}(x,y)_{\rm sp}^2}{2}+\mathcal{O}(x^9)\ ,
\end{equation}
a perfect agreement up to the order we checked.

\section{The index for M2-branes probing $C(M^{32})$}

\subsection{Gravity}

The gravity spectrum of M-theory on $AdS_4\times M^{32}$ was obtained in
\cite{Fabbri:1999mk}. Among them, in this paper we are only interested in the BPS states
saturating the bound $\epsilon\geq h_0+j_3$, where $h_0$ is the $U(1)$ R-charge. Exactly
one field in $AdS_4$ appears to be BPS in each short $OSp(2|4)$ supermultiplet. Those fields
are listed in Table \ref{M32-KK} with their quantum numbers for energy $\epsilon$, angular
momentum $j_3$, $SU(3)$ label $(M_1,M_2)$ which follows the same convention as \cite{Fre':1999xp}
and our presentation for $N^{010}$ in section 2, $SU(2)$ global symmetry. The Cartan of $SU(2)$
global symmetry is identified as a `baryon-like' $U(1)$ symmetry carried by the
sum of monopole charge $\sum p_i$ in the first gauge group $U(N)_{-2}$. This $U(1)$ which
we often call `baryon-like' should not be confused with the real baryon $U(1)$ symmetry,
as $M^{32}$ has topological 5-cycles on which M5-brane baryons can wrap.
\begin{table}[t!]
$$
\begin{array}{c|ccccc}
  \hline &\epsilon&j_3&\epsilon+j_3&SU(3)&SU(2)\\
  \hline p\geq 1&2p+5/2&3/2&2p+4&(0,2p)&p\\
  \hline p\geq 0&2p+3&1&2p+4&(1,3p+1)&p+1\\p\geq 1&2p+1&1&2p+2&(0,3p)&p-1\\
  \hline p\geq 1&2p+3/2&1/2&2p+2&(1,3p+1)&p\\
  p\geq 1&2p+3/2&1/2&2p+2&(0,3p)&p+1\\
  \hline p\geq 1&2p&0&2p&(0,3p)&p\\ \hline\cdot&5/2&3/2&4&(0,0)&0\\
  \cdot&3/2&1/2&2&(1,1)&0\\ \cdot&3/2&1/2&2&(0,0)&1\\
  \cdot&3/2&1/2&2&(0,0)&0\\
  \hline
\end{array}
$$
\caption{Supersymmetric fields on $AdS_4\times M^{32}$}\label{M32-KK}
\end{table}
In the table, the first line comes from short graviton multiplets, the second and third lines
from the short gravitino multiplets, the fourth and fifth lines from the short vector multiplets,
sixth from the hypermultiplets. The seventh line is from the massless graviton multiplets.
The eighth and ninth line is from the massless vector multiplet for the $SU(3)$ and $SU(2)$
isometry, respectively. The last line is from the $U(1)$ baryon symmetry for M5-branes wrapped
on 5-cycles in $M^{32}$.

Each field on the table contributes to the index. To calculate the single particle index,
one also has to take the wavefunction factors in $AdS_4$ into account. This amounts to putting
the factor $\frac{1}{1-x^2}$ to the index over the above fields.
Denoting by $h_1,h_2$ the Cartans of $SU(3)$ given by the diagonal matrices
$\frac{1}{2}{\rm diag}(1,-1,0)$, $\frac{1}{2}{\rm diag}(0,1,-1)$ and by $h_3$ the $SU(2)$
Cartan, one obtains
\begin{eqnarray}
  I_{\rm sp}(x,y_1,y_2,y_3)&=&{\rm tr}\left[(-1)^Fx^{\epsilon+j_3}y_1^{h_1}y_2^{h_2}y_3^{h_3}
  \right]\\
  &=&\frac{1}{1-x^2}\left[-\sum_{p=1}^\infty x^{2p+4}\chi^{(0,2p)}_{SU(3)}(y_1,y_2)
  \chi_{SU(2)}^p(y_3)+\right]\nonumber
\end{eqnarray}
where
\begin{equation}
  \chi_{SU(3)}^{M_1,M_2}(y_1,y_2)=\frac{\left|\begin{array}{ccc}
  y_1^{M_1\!+\!M_2\!+\!2}&1/y_2^{M_1\!+\!M_2\!+\!2}&(y_2/y_1)^{M_1\!+\!M_2\!+\!2}\\
  y_1^{M_2\!+\!1}&1/y_2^{M_2\!+\!1}&(y_2/y_1)^{M_2\!+\!1}\\1&1&1\end{array}\right|}
  {\left|\begin{array}{ccc}y_1^2&1/y_2^2&y_2^2/y_1^2\\y_1&1/y_2&y_2/y_1\\1&1&1\end{array}\right|}
\end{equation}
and
\begin{equation}
  \chi_{SU(2)}^j(y_3)=\frac{y_3^{j\!+\!1/2}-y_3^{-(j\!+\!1/2)}}{y_3^{1/2}-y_3^{-1/2}}
\end{equation}
are the $SU(3)$ and $SU(2)$ characters.

The general expression for $I_{\rm sp}(x,y_1,y_2,y_3)$ after the infinite sums is very messy,
although an explicit closed form expression is available. The expressions we need below
to compare with field theory would be those with definite $SU(2)$ Cartan charge, labeled by
$y_3$. This charge would correspond to the $U(1)_B$ baryon-like charge carried by the monopoles.
The single particle indices with this charge being $0$, $1$, $2$ are
\begin{eqnarray}\label{gravity-M32}
  I^{(0)}_{\rm sp}&=&\frac{x^2/y_1^3}{1-x^2/y_1^3}+\frac{x^2y_2^3}{1-x^2y_2^3}+
  \frac{x^2y_1^3/y_2^3}{1-x^2y_1^3/y_2^3}-\frac{3x^2}{1-x^2}\\
  I^{(1)}_{\rm sp}&=&\frac{x^2/y_1^3}{1-x^2/y_1^3}+\frac{x^2y_2^3}{1-x^2y_2^3}+\frac{x^2y_1^3/y_2^3}
  {1-x^2y_1^3/y_2^3}-\frac{x^4}{1-x^2}\left(1/y_1^3+y_2^3+y_1^3/y_2^3\right)\nonumber\\
  &&+x^2\left(y_1^2/y_2+y_2^2/y_1+y_1/y_2^2+y_2/y_1^2+y_1y_2+1/(y_1y_2)\right)\nonumber\\
  I^{(2)}_{\rm sp}&=&
  \frac{(x^8+x^4y_1^3-2x^6y_1^3+x^2y_1^6-x^4y_1^6-y_1^9+x^2y_1^9-x^6y_1^9+x^8y_1^9+x^4y_1^{12}-
  x^6y_1^{12})}{(-1+x^2)y_1^6(x^2-y_1^3)}\nonumber\\
  &&+\frac{x^6y_1^6}{(-1+x^2)y_2^6}+\frac{x^4y_1^4}{y_2^5}+\frac{x^4y_1^2+x^4y_1^5}{y_2^4}
  +\frac{x^4-x^2y_1^3}{y_2^3}+\frac{x^4+x^4y_1^6}{y_1^2y_2^2}+\frac{x^4}{y_1^4y_2}+\frac{x^4y_2}{y_1^5}
  +\frac{x^4(1+y_1^6)y_2^2}{y_1^4}\nonumber\\
  &&+\frac{x^2(x^2-y_1^3)y_2^3}{y_1^3}+\frac{x^4(1+y_1^3)y_2^4}{y_1^2}
  +\frac{x^4y_2^5}{y_1}+\frac{x^6y_2^6}{-1 + x^2}+\frac{x^2 y_1^3}{(-x^2 y1^3 + y2^3)}-\frac{1}{(-1 + x^2 y_2^3)}\nonumber
\end{eqnarray}
and so on, where $I^{(p)}_{\rm sp}=\oint_{|y_3|=1}\frac{dy_3}{y_3^{p\!+\!1}}I_{\rm sp}(x,y_1,y_2,y_3)$
is the single particle index with $U(1)_B$ charge $p$. Note that, as one takes the $SU(3)$
chemical potentials $y_1,y_2$ to $1$, a dramatic simplification (or cancelation) appears:
\begin{equation}
  I_{\rm sp}(x,y_1\!=\!1,y_2\!=\!1,y_3)=\frac{9x^2y_3}{(1-x^2y_3)^2}+\frac{9x^2y_3^{-1}}
  {(1-x^2y_3^{-1})^2}\ ,
\end{equation}
or
\begin{equation}
  I^{(0)}_{\rm sp}=0\ ,\ \ I_{\rm sp}^{(1)}=9x^2\ ,\ \ I^{(2)}_{\rm sp}=18x^4\ ,\ \
  I^{(3)}_{\rm sp}=27x^6\ ,\ \cdots
\end{equation}
and so on.

\subsection{Field theory}

\begin{figure}[t!]
  \begin{center}
    \includegraphics[width=6cm]{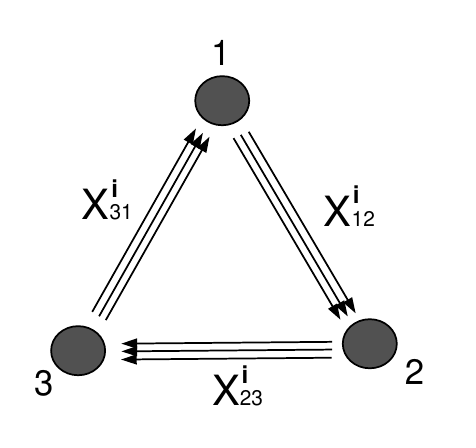}
\caption{The quiver diagram of M2-branes on $C(M^{32})$}\label{quiver-m32}
  \end{center}
\end{figure}

The field theory proposed for M2-branes on $C(M^{32})$ is given by the quiver diagram
in Fig \ref{quiver-m32}. The Chern-Simons levels for the three $U(N)$ gauge fields are
$(-2k,k,k)$ for $M^{32}/\mathbb{Z}_k$ \cite{Martelli:2008si,Hanany:2008cd}.
For the brevity of notation, we shall often denote the fields by $X^i_{23}=A^i$,
$X^i_{31}=B^i$, $X^i_{12}=C^i$. $i=1,2,3$ is the triplet index for the $SU(3)$ global
symmetry.

This $\mathcal{N}\!=\!2$ model for $M^{32}$ is chiral in a 4 dimensional sense.
As the number of `flavors' $3$ for each bi-fundamental is odd, one has to worry about
the issue of parity anomaly. The fermion determinant could pick up a minus sign under certain
large gauge transformations,
which will make the path integral gauge non-invariant.  More concretely, on $S^2\times S^1$
that we are going to consider, the relevant large gauge transformations are $2\pi$ periodic
shifts of the holonomy variables in our convention. The possible $-1$ sign will
come from the phase $e^{ib_0}$ defined below.

The canonical way of making such a theory consistent would be adding Chern-Simons term
with appropriate half-integral levels, providing an extra minus sign to the path integral
measure under the large gauge transformation which cancels that from the fermion determinant.
As we shall see in the context of index shortly, the required Chern-Simons term is a mixed
Chern-Simons term between the three gauge fields. As we shall see in the context of the index
below, the required mixed Chern-Simons term is needed only for the overall $U(1)^3$ part of
the $U(N)^3$ gauge fields. It should be very interesting to study the index for a suitably
modified theory of this sort.\footnote{We thank I. Klebanov for discussions,
and especially F. Benini for explaining to us his studies on this issue.}

In this section, we try to do best to make the proposed model of \cite{Martelli:2008si,Hanany:2008cd}
sensible in the context of the superconformal index. To this end, let us consider at this point
another possible way of making the path integral consistent on $S^2\times S^1$,
which later will turn out to be a prescription yielding an interesting index structure
containing all the gravity spectrum. At least when one considers a theory on $S^2\times\mathbb{R}$
(or $S^2\times S^1$) rather than generic case, there appear nontrivial sectors
labeled by the magnetic monopole charges. Depending on these magnetic charges that we allow
in this QFT as we shall explain below, it is possible to make the path integral measure
to be invariant under the large gauge transformations of holonomies. We postpone
the explanation until we have a concrete form of this measure.

We take the dimensions of three chiral fields $A^i$, $B^i$, $C^i$ to be $a$, $b$, $c$,
respectively. They should satisfy $a+b+c=2$ for the superpotential to be marginal.
In principle, the parameters $a,b,c$, including the superpotential constaint, could
be derivable from an extremization principle like that of \cite{Jafferis2}. We simply start
with them unfixed, but at appropriate stage will constrain them by hand if necessary, to have
a sensible model with the $M^{32}$ gravity dual. The only other constaint we impose will
be $b=c$, which is anyway reasonable from the quiver diagram.

From the general expression in \cite{Imamura:2011su}, one can immediately write down
the integral expression for the index for this model:
\begin{eqnarray}\label{M32-index}
  \hspace*{-1cm}I_{(p,q,r)}&=&\frac{x^{\epsilon_0}}{({\rm symmetry})}\int\left[
  \frac{d\alpha d\beta d\gamma}{(2\pi)^3}\right]e^{ik\sum_i(-2p_i\alpha_i+q_i\beta_i+r_i\gamma_i)}
  e^{ib_0(p,q,r,\alpha,\beta,\gamma)}\\
  \hspace*{-1cm}&&\exp\left[\frac{}{}\!\right.\sum_{i,j=1}^N\sum_{n=1}^\infty\frac{1}{n}
  x^{n|q_i-r_j|}\left(f_A^+(\cdot^n)e^{-in(\beta_i-\gamma_j)}+f_A^-e^{in(\beta_i-\gamma_j)}\right)
  +(A,\beta,\gamma,q,r\rightarrow B,\gamma,\alpha,r,p)\nonumber\\
  &&+(A,\beta,\gamma,q,r\rightarrow C,\alpha,\beta,p,q)\left.\frac{}{}\!-\sum_{i\neq j}
  \sum_{n=1}^\infty\frac{1}{n}x^{n|p_i-p_j|}e^{-in(\alpha_i-\alpha_j)}+
  (p,\alpha\rightarrow q,\beta)+(p,\alpha\rightarrow r,\gamma)\right]\nonumber
\end{eqnarray}
with
\begin{equation}
  f_A^+=\frac{x^a}{1-x^2}\left(1/y_1+y_2+y_1/y_2\right)\ ,\ \ f_A^-=-\frac{x^{2-a}}{1-x^2}
  \left(y_1+1/y_2+y_2/y_1\right)\ ;\ \
  f_{B,C}^\pm=({\rm replace}\ a\rightarrow b,c)
\end{equation}
and
\begin{eqnarray}
  \hspace*{-0.5cm}b_0&=&\frac{3}{2}\left[\sum|p_i-q_j|(\alpha_i-\beta_j)+
  \sum|q_i-r_j|(\beta_i-\gamma_j)+\sum|r_i-p_j|(\gamma_i-\alpha_j)\right]\\
  \hspace*{-0.5cm}\epsilon_0&=&\frac{3}{2}\left((1\!-\!c)\!\sum_{i,j}|p_i\!-\!q_j|+(1\!-\!a)
  \!\sum_{i,j}|q_i\!-\!r_j|+(1\!-\!b)\!\sum_{i,j}|r_i\!-\!p_j|\right)-\sum_{i<j}\left(|p_i\!-\!p_j|+
  |q_i\!-\!q_j|+|r_i\!-\!r_j|\right)\ .\nonumber
\end{eqnarray}
The set of $N$ integers $\{p_i\}$, $\{q_i\}$, $\{r_i\}$ are monopole charges in $U(N)_{-2k}$,
$U(N)_k$, $U(N)_k$ gauge groups. We shall be mostly interested in the case with $k=1$ below.

Let us make comment on the 1-loop shift of the phase $b_0$, which is
the index version of the 1-loop fermion determinant contributing to
the Chern-Simons term. This could pick up dangerous minus signs by
the large gauge transformations. The large gauge transformations in
this case are periodic shifts of the holonomy variables $\alpha_i$,
$\beta_i$, $\gamma_i$ by $2\pi$, making them live on a $3N$
dimensional torus. With generic integers $\{p_i\}$, $\{q_i\}$,
$\{r_i\}$, the phase $e^{ib_0}$ may not be well-defined on this
torus due to the $\frac{3}{2}$ factor. However, by suitably
restricting these integers, constraining the monopole sectors of the
theory, we can make this phase well defined. Let us see how this can
happen.

Let us first consider the appearance of a variable $\alpha_i$ in the phase $e^{ib_0}$:
\begin{equation}
  \exp\left[\frac{3}{2}\alpha_i\sum_{j=1}^N\left(|p_i-q_j|-|p_i-r_j|\right)\right]
  =\exp\left[\frac{3}{2}\alpha_i\sum_{j=1}^N\left(|p_i|+|q_j|-|p_i|-|r_j|\right)\right]
  e^{\frac{3}{2}\alpha_i({\rm even\ ingeter})}\ ,
\end{equation}
where we used the fact that $|m|+|n|-|m-n|$ is an even integer for integer $m,n$. The
second phase in the last expression is thus well defined. To have the first part well defined,
we demand the restriction
\begin{equation}
  \sum_{i=1}^N|q_i|=\sum_{i=1}^N|r_i|\ \ \mod 2\ .
\end{equation}
Considering the large gauge transformation of $\beta_i$, $\gamma_i$, one finds that
the restriction is $\sum|p|=\sum|q|=\sum|r|$ mod $2$. This can again be written as
\begin{equation}
  \sum_{i=1}^N p_i=\sum_{i=1}^Nq_i=\sum_{i=1}^Nr_i\ \ \mod 2\ .
\end{equation}
These are actually constraining the overall $U(1)^3$ fluxes in $U(N)^3$,
\begin{equation}
  \int_{S^2}{\rm tr}F_1=\int_{S^2}{\rm tr}F_2=\int_{S^2}{\rm tr}F_3\ \ \mod 2\ ,
\end{equation}
which are $U(N)^3$ gauge invariant constraints. One interpretation is that this condition
comes from different quantization for Chern-Simons level for $U(1)$ of $U(N)$.
Assigning different Chern-Simons level for $U(1)$ part and $SU(N)$ part
out of $U(N)$ is possible. This suggests that we have to choose the different Chern-Simons levels
for $U(1)$ and $SU(N)$  to make theory free of the potential parity anomalies due to fermion one-loop
determinants. It would be interesting to see if a similar consideration works
for other manifolds, say on $S^3$. Below, we simply consider the
index (\ref{M32-index}) with this restriction understood.

Again our main concern of this paper is the large $N$ index with energies kept at $\mathcal{O}(1)$.
In particular, we again require only $\mathcal{O}(1)$ number of magnetic charges to be nonzero
among the above $3N$ of them. The details of the large $N$ integration over holonomies with zero
flux is analogous to the previous section. After this calculation, the large $N$ index is given by
\begin{equation}
  I_{N=\infty}(x,y_1,y_2,y_3)=I_0(x,y_1,y_2)I^\prime(x,y_1,y_2,y_3)\ ,
\end{equation}
where
\begin{equation}\label{I0-M32}
  I_0=\prod_{n=1}^\infty\frac{(1-x^{2n})^3}{(1-x^{2n}/y_1^{3n})(1-x^{2n}y_2^{3n})
  (1-x^{2n}y_1^{3n}/y_2^{3n})}\equiv\exp\left[\sum_{n=1}^\infty\frac{1}{n}
  f(x^n,y_1^n,y_2^n)\right]
\end{equation}
with
\begin{equation}
  f(x,y_1,y_2)=\frac{x^2/y_1^3}{1-x^2/y_1^3}+\frac{x^2y_2^3}{1-x^2y_2^3}+
  \frac{x^2y_1^3/y_2^3}{1-x^2y_1^3/y_2^3}-\frac{3x^2}{1-x^2}\ .
\end{equation}
The monopole part $I^\prime$ is given by
\begin{eqnarray}
  \hspace*{-2cm}I^\prime_{(p,q,r)}&=&\frac{x^{\epsilon_0}}{({\rm symmetry})}\int\left[\frac{d\alpha d\beta d\gamma}{(2\pi)^3}\right]e^{ik\sum_i(-2p_i\alpha_i+q_i\beta_i+r_i\gamma_i)}
  e^{ib_0(p,q,r,\alpha,\beta,\gamma)}\\
  \hspace*{-2cm}&&\hspace{-2cm}\exp\left[\frac{}{}\!\right.\sum_{i,j=1}\sum_{n=1}^\infty\frac{1}{n}
  (x^{n|q_i-r_j|}-x^{n|q_i|+n|r_j|})\left(f_A^+(\cdot^n)e^{-in(\beta_i-\gamma_j)}+f_A^-e^{in(\beta_i-\gamma_j)}\right)
  +(A,\beta,\gamma,q,r\rightarrow B,\gamma,\alpha,r,p)\nonumber\\
  \hspace*{-2cm}&&\hspace{-2cm}+(A,\beta,\gamma,q,r\rightarrow C,\alpha,\beta,p,q)-\sum_{i,j}
  \sum_{n=1}^\infty\frac{1}{n}\left((1-\delta_{ij})x^{n|p_i-p_j|}-x^{n|p_i|+n|p_i|}\right)
  e^{-in(\alpha_i-\alpha_j)}+(p,\alpha\rightarrow q,\beta)+(p,\alpha\rightarrow
  r,\gamma)\left.\frac{}{}\!\right]\nonumber
\end{eqnarray}
where the holonomies $\alpha_i,\beta_i,\gamma_i$ range over those supporting nonzero fluxes.
For a given set of fluxes, this index does not have any factorization structure.

As the diagonal combination of $U(1)\subset U(N)_{-2k}\times U(N)\times U(N)_k$ decouples with
matter fields, the fluxes should satisfy
\begin{equation}\label{overall-decouple}
  2\sum p_i\!=\!\sum q_i+\sum r_i\ .
\end{equation}
Also, the magnetic charge
\begin{equation}
  \sum_ip_i
\end{equation}
for the first gauge group $U(N)_{-2k}$ is to be identified with the Cartan of the $SU(2)$
flavor symmetry at $k\!=\!1$, which is explicit in the gravity side. We call this the $U(1)_B$
`baryon-like' symmetry, again not to be confused withe the real baryon symmetry for
wrapped M5-branes.

As our first goal is to study gravitons with $\mathcal{O}(1)$ numbers of magnetic flux,
we consider the condition for the energy to remain $\mathcal{O}(1)$. Firstly, from the phase
$b_0$, it happens that there can appear $O(N)$ number of phase variables in the exponent
even with $\mathcal{O}(1)$ number of fluxes. This would cause gauge invariant operators to
carry $\mathcal{O}(N)$ energy after integration. For this not to happen, one finds that
\begin{equation}\label{O(1)-energy}
  \sum|p|=\sum|q|=\sum|r|
\end{equation}
has to be imposed, where the summations are over all nonzero fluxes in each gauge group.
Of course this is compatible with the flux restriction we imposed above.
Let us next investigate the possible $\mathcal{O}(N)$ contribution to $\epsilon_0$. Denoting
by $N_1,N_2,N_3$ the number of zero magnetic fluxes in each gauge group, one finds
\begin{equation}\label{O(N)-energy}
  N_1\left(\frac{3}{2}(1-c)\sum|q|+\frac{3}{2}(1-b)\sum|r|-\sum|p|\right)+cyclic.
\end{equation}
When the condition (\ref{O(1)-energy}) is met, the $\mathcal{O}(N)$
contribution always vanishes from $a+b+c=2$.

Firstly, the expression (\ref{I0-M32}) for $I_0$ from field theory completely agrees with
the gravity index $I^{(0)}$ coming from $U(1)_B$ neutral particles, similar to the
$\mathcal{N}\!=\!6$ models and also the models studied in the previous sections. Actually,
when the large $N$ index is factorized into $I_0I^+I^-$ acquiring
contribution from $U(1)_B$ neutal sector, positive monopoles and negative monopoles,
respectively, this agreement $I_0=I^{(0)}$ has to be the case. But as we shall see more concretely
with examples, the index $I^\prime$ does not generally factorize to $I^+I^-$ for separate
saddle points (i.e. monopole fluxes). Below, we shall explain how such a factorization can
appear \textit{after} summing over the saddle points.

Let us first phrase our claim, and then illustrate by many examples. For a given saddle point,
we decompose nonzero magnetic charges $\{p_i\}$ into positive and negative ones, and call them
$p_{i+}$, $p_{i-}$, respectively. We will set $b=c$ for the trial R-charge, which we shall motivate
shortly with examples, so that the only independent parameter left is $b$. Define
\begin{equation}
  P_+=\sum p_{i+}\ ,\ \ P_-=\sum |p_{i-}|
\end{equation}
to be the sums of all positive/negative fluxes. Defining $I_{(P_+,P_-)}$ to be the sum over all
indices whose saddle points have given values of $P_+$, $P_-$, this quantity can be written as
\begin{equation}\label{M32-decompose}
  I_{(P_+,P_-)}=\hat{I}_{(P_+,P_-)}+I^b_{(P_+,P_-)}\ ,
\end{equation}
where $\hat{I}$ refers to the index summed over the sectors whose energies do not depend on $b$,
and $I^b$ that does depend on $b$.\footnote{Coefficients of the expansion with chemical potentials
are always independent of $b$ in both parts, being integers from the Plethystic exponential structure.}
We find that
\begin{equation}\label{M32-factorize}
  \hat{I}_{(P_+,P_-)}=\hat{I}_{(P_+,0)}\hat{I}_{(0,P_-)}=I_{P_+}^{grav}I_{P_-}^{grav}\ ,
\end{equation}
i.e. $\hat{I}$ factorizes into two contribution which are identical to the multi-graviton
index $I^{grav}_{P_+}$ with $U(1)_B$ charge $P_+$ coming from positively charged particles, and
the index $I^{grac}_{P_-}$ with charge $-P_-$ coming from negatively charge particles. If, for some
reason, one has to discard all the seemingly existing saddle points contributing to $I^b$,
then one finds from (\ref{M32-factorize}) that the following factorization is allowed.
Labeling the index with its $U(1)_B$ charge $P_+-P_-$ with the chemical potential $y_3$,
the above factorization implies
\begin{eqnarray}
  \hat{I}^\prime(x,y_1,y_2,y_3)&=&\sum_{P_+,P_-=0}^\infty y_3^{P_+-P_-}\hat{I}_{(P_+,P_-)}(x,y_1,y_2)\\
  &=&\sum_{P_+=0}^\infty y_3^{P_+}\hat{I}_{(P_+,0)}(x,y_1,y_2)\sum_{P_-=0}^\infty
  y_3^{-P_-}\hat{I}_{(0,P_-)}(x,y_1,y_2)\equiv I^+I^-\ .
\end{eqnarray}
$I^+$ and $I^-$ separately agrees with the gravity index coming from particles with
positive/negative charges.

Before proceeding, let us explain what could be the meaning of discarding/keeping $I^b$.
We have two possibilities in mind. Firstly, quite naturally, if we seriously keep this
contribution, this could simply be reflecting the fact that the proposed field theory
is not the correct one for M-theory on $AdS_4\times M^{32}$, as the index captures extra
branches of states other than those seen in the gravity. This looks a bit similar to
the phenomenon observed in the so-called `dual ABJM' model \cite{Hanany:2008fj} that
extra branches of moduli spaces develop \cite{RodriguezGomez:2009ae}, although appearing
here in a much more nontrivial way of showing two sectors in the index.
As we shall explain in more detail below, the fluxes contributing to $\hat{I}$ and $I^b$
have clear distinctions in their properties. So even though this model itself does not
seem to be showing the correct gravity spectrum, it still sounds quite amazing that the
known gravity contribution is so well organized as above. So we could hope that a small
modification of the existing model could provide a much better index structure.

One the other hand, let us also emphasize a potential subtlety of the localization
calculation in quantum field theories which may justify discarding $I^b$ as unphysical.
Localization refers to a property of an integral with
fermionic, or nilpotent, symmetry $Q$ which makes the integral to acquire contribution
only around the fixed points of the symmetry. In particular, a practical way of carrying
out such integrals is to add to the integrand $Q$-exact measures with free parameters,
using which one can perform a Gaussian evaluation of the exact result. The location of the
fixed points of the symmetry in the integration domain could change during this deformation.
This change of the fixed point profiles does not change the result \textit{as long as the
saddle points do not disappear or appear} during the deformation. This can happen if the
integration domain is noncompact. A classic example of this sort appears in the calculation
of the partition function of 2 dimensional topological Yang-Mills theory \cite{Witten:1992xu}.

The localization calculation of the index, or the path integral for the partition function of
Chern-Simons-matter theory on $S^2\times S^1$ in principle has the same issue when we introduce
the Yang-Mills like deformation of \cite{Kim:2009wb}. It is not easy to determine a priori which
saddle points are to be kept and which to be discarded. In the case of $\mathcal{N}\!=\!6$ theory,
originally some saddle points which appeared in the deformed theory of \cite{Kim:2009wb}
were not clearly understood in the QFT without deformation. Many of these saddle points
were constructed later from the undeformed theory \cite{Kim:2010ac}, justifying in the honest
sense the prescription of \cite{Kim:2009wb}.

In the present case of $M^{32}$, the observation summarized above
suggest that all saddle points which contain the undetermined trial
R-charge parameter $b$ might have to be discarded as being
unphysical ones. Also, the possibility of discarding $I^b$ is not
simply discarding all unwanted terms in the index, as the ambiguity
mentioned above only allows one to keep or discard the whole
contribution from a saddle point, not allowing to keep some term
while discarding others. It should be interesting to see if this is
the case or not for this model, using the QFT on
$S^2\times\mathbb{R}$ written down in \cite{Imamura:2011su} with and
without deformations.

Now let us present examples which shows that our claims (\ref{M32-decompose}), (\ref{M32-factorize})
are true in various sectors. The sector with $p=q=r=1$ is for smallest positive $U(1)_B$ charge.
Inserting $\epsilon_0=0$ and $b_0=0$, the field theory index is
\begin{equation}\label{(1)(1)(1)}
  \int_{\alpha,\beta,\gamma}e^{i(-2\alpha+\beta+\gamma)}
  \frac{(1-x^{2-c}e^{i(\alpha-\beta)})^3(1-x^{2-a}e^{i(\beta-\gamma)})^3(1-x^{2-b}e^{i(\gamma-\alpha)})^3}
  {(1-x^ce^{-i(\alpha-\beta)})^3(1-x^ae^{-i(\beta-\gamma)})^3(1-x^be^{-i(\gamma-\alpha)})^3(1-x^2)^3}\ .
\end{equation}
The result turns out to be extremely simple: $9x^{a+2b}$. Note that in this simple result, even
the constant infinite series factor $\frac{1}{(1-x^2)^3}$ is canceled out, implying that there
should be a large amount of cancelation between bosonic and fermionic operators.

Let us explain the first few terms to illustrate how to understand this number, and
how the higher order cancelation is happening. We should suitably excite
the fields to obtain the phase $e^{i(2\alpha-\beta-\gamma)}\equiv\frac{u^2}{vw}$. There are two
possible ways of obtaining it at the lowest energy $x^{a+2b}$. This is either by exciting scalars $A_iB_{(j}B_{k)}$ carrying $\frac{w}{v}\left(\frac{u}{w}\right)^2$ phase which yields $(3x^a)(6x^{2b})$,
or by exciting $B_i\bar\Psi_{Cj}$ as $\frac{u}{v}\frac{u}{w}$ with $(3x^b)(-3x^{2-c})$, yielding
$(28-9)x^{a+2b}=9x^{a+2b}$ in total and explains the above result.
One may further wonder what happens at higher orders.
One can easily find that the phase can also be provided as $\left(\frac{u}{v}\right)^2\frac{v}{w}$
by $\psi_{C[i}\psi_{Cj]}\psi_{Ak}$ with the index $(3x^{2a+2b})(-3x^{b+c})=-9x^{a+2b+2}$,
and  by $(AB^2-B\psi_C)(ABC-B\psi_B)$ with $(6\cdot 10\cdot 3-3^3\cdot 6-10\cdot 3^2+6\cdot 3^2)x^{a+2b+2}
=-18x^{a+2b+2}$. So at this order, one obtains $9x^{a+2b}(1-x^2)^3$, where the second factor cancels
the overall factor $\frac{1}{(1-x^2)^3}$, consistent with our general computation.

We should identify $9x^{a+2b}$ with suitable gravity states, which will give us
a hint about the values of the R-charge $a$, $b$. Obviously, there is a cancelation $18-9=9$
between bosonic and fermionic modes. As for the bosonic modes $AB^2$, it is easy to guess what
are the true BPS operators, as they are chiral rings. One simply symmetrizes all three $SU(3)$
indices, obtaining $10x^{a+2b}$. So this should be part of the hypermultiplet with $p\!=\!1$
and SU(3) representation $(0,3)$. Thus, we find $a+2b=2$, i.e. $b\!=\!c$. This in turn implies
that only one out of $9$ operators $B\psi_C$ survive to be BPS. It is easy to find what they are
in the gravity dual. They are the fermions in the massless vector multiplet for $SU(2)$ symmetry,
as the operator carries nonzero $U(1)_B$ to be the $SU(2)$ Cartan. Therefore, only the $SU(3)$
singlet survives to be BPS.

In fact, summing over an infinite tower of all single gravity particles with $U(1)_B$ charge
$1$, one obtains $I^{(1)}_{\rm sp}(x)=9x^2$ as shown in the previous subsection, agreeing
with $I_{(1)(1)(1)}$ calculated above. More generally, with all chemical potentials $y_1,y_2$
kept, we have also checked that the integral (\ref{(1)(1)(1)}) again
completely reproduces the gravity index $I{(1)}_{\rm sp}$ of (\ref{gravity-M32}).

To get a more nontrivial test of our claim,
we investigate the sector with charge $2$ with $P_+=2$ and $P_-=0$. We shall compare
the index with the charge $2$ gravity index made of positively charge particles only.
Up to $\mathcal{O}(x^{12})$ and for any trial R-charge $b$, we find
\begin{eqnarray}
  I_{(2)(2)(2)}&=&18x^4+9x^8-27x^{10}-27x^{12}\cdots\nonumber\\
  I_{(1,1)(1,1)(1,1)}&=&45x^4-54x^8+378x^{10}-1053x^{12}\cdots\nonumber\\
  I_{(1,1)(1,1)(2)}&=&0+\cdots\nonumber\\
  I_{(1,1)(2)(1,1)}&=&45x^8-360x^{10}+1251x^{12}\cdots\nonumber\\
  I_{(2)(1,1)(1,1)}&=&0+
  \cdots\nonumber\\
  I_{(2)(2)(1,1)}&=&9x^8-18x^{10}-171x^{12}\cdots\nonumber\\
  I_{(2)(1,1)(2)}&=&-9x^8+27x^{10}+27x^{12}\cdots\nonumber\\
  I_{(1,1)(2)(2)}&=&-27x^{12}\cdots\ .
\end{eqnarray}
At $\mathcal{O}(x^4)$, the first two contributions are the expected ones. The first is
the single particle contribution with $U(1)_B$ charge $2$, from hypermultiplets scalars
in $(0,6)$ and vector multiplet fermion in $(0,3)$: $28-10-18$. The second is simply the
2-particle states of the $9$ charge $1$ particles: $9+36=45$. At higher orders,
the sum of the above indices is $63x^4-\mathcal{O}(x^{14})$.

From the results summarized in the previous subsection,
the single particle gravity index at charge $2$ and 2 particle index
of charge $1$ particles are given by
\begin{equation}
  I^{(2)}_{\rm sp}(x)=18x^4\ ,\ \ \frac{I^{(1)}_{\rm sp}(x^2)+I^{(1)}_{\rm sp}(x)^2}{2}
  =45x^4\ ,
\end{equation}
again after a vast cancelation of the infinite tower of graviton states, similar to the
charge $1$ single particle index. Although the expression is too messy to be recorded here,
we report that we have also checked the agreement after keeping the $SU(3)$ chemical
potentials $y_1,y_2$. Therefore, the positive flux part of the index agrees with the
charge $2$ field theory index restricted to the positive fluxes up to $\mathcal{O}(x^{14})$,
and we claim this happens to all orders.

To study more nontrivial sectors, we study the sector with $U(1)_B$ charge $1$
with $P_+=2$, $P_-=1$. To classify the fluxes in this sector, we denote by
$P_\pm\equiv\sum|p_\pm|$, $O_\pm\equiv\sum|q_\pm|$,
$R_\pm\equiv\sum|r_\pm|$ the sum over positive/negative fluxes in a gauge group. The
conditions for decoupling overall $U(1)$ and $\mathcal{O}(1)$ energy are
\begin{equation}
  P_++P_-=Q_++Q_-=R_++R_-\ ,\ \ Q_+-Q_-+R_+-R_-=2(P_+-P_-)=2\ .
\end{equation}
These fluxes are again classified by a non-negative integer $n$, with $P_+\!=\!n\!+\!1$,
$P_-\!=\!n$. As the sector labeled by $n$ starts at the order $x^{4n\!+\!2}$, we study
the saddle points with $P_+=2,P_-=1$ which corrects the $n\!=\!0$ sector (positive flux only)
that we considered above. Possible values of flux sums are
\begin{equation}
  (Q_+,Q_-,R_+,R_-)=(2,1,2,1),\ (1,2,3,0),\ (3,0,1,2)\ .
\end{equation}
The fluxes in the first case and the correspondinc indices are as follows:
\begin{eqnarray}
  I_{(2,-1)(2,-1)(2,-1)}&=&I_{(2)(2)(2)}I_{(-1)(-1)(-1)}=9x^2(18x^4+9x^8-27x^{10}-27x^{12}+\cdots)\nonumber\\
  I_{(1,1,-1)(1,1,-1)(1,1,-1)}&=&I_{(1,1)(1,1)(1,1)}I_{(-1)(-1)(-1)}=9x^2
  (45x^4-54x^8+378x^{10}-1053x^{12}\cdots)\nonumber
\end{eqnarray}
\begin{eqnarray}
  I_{(2,-1)(2,-1)(1,1,-1)}&=&I_{(2)(2)(1,1)}I_{(-1)(-1)(-1)}=9x^2(9x^8-18x^{10}-171x^{12}\cdots)\nonumber\\
  I_{(2,-1)(1,1,-1)(2,-1)}&=&I_{(2)(1,1)(2)}I_{(-1)(-1)(-1)}=9x^2(-9x^8+27x^{10}+27x^{12}\cdots)\nonumber\\
  I_{(1,1,-1)(2,-1)(2,-1)}&=&I_{(1,1)(2)(2)}I_{(-1)(-1)(-1)}=9x^2(-27x^{12}+\cdots)\nonumber
\end{eqnarray}
\begin{eqnarray}
  I_{(2,-1)(1,1,-1)(1,1,-1)}&=&162 (27x^{16}-125x^{18}+169x^{20}+\cdots)\nonumber\\
  I_{(1,1,-1)(2,-1)(1,1,-1)}&:&405x^{10}-3240x^{12}+11259x^{14}+\cdots\nonumber\\
  I_{(1,1,-1)(1,1,-1)(2,-1)}&:&0x^{14}+\cdots\ .
\end{eqnarray}
The $12$ fluxes in the second case are
\begin{eqnarray}
  I_{(2,-1)(1,-2)(3)}&=&x^{4b}\left(9x^6-9x^8-171x^{10}-135x^{12}+747x^{14}+
  1323x^{16}+\cdots\right)\nonumber\\
  I_{(2,-1)(1,-2)(2,1)}&=&x^{4b}\left(-63x^8+144x^{10}+270x^{12}-351x^{14}-1440x^{16}+\cdots\right)\nonumber\\
  I_{(2,-1)(1,-2)(1,1,1)}&=&x^{4b}\left(180x^{14}-261x^{16}+\cdots\right)\nonumber\\
  I_{(2,-1)(1,-1,-1)(3)}&=&x^{4b}\left(-x^{10}+x^{12}+17x^{14}+17x^{16}+\cdots\right)\nonumber\\
  I_{(2,-1)(1,-1,-1)(2,1)}&=&x^{4b}\left(126x^{12}-288x^{14}-288x^{16}+\cdots\right)\nonumber\\
  I_{(2,-1)(1,-1,-1)(1,1,1)}&=&x^{4b}\left(0x^{16}+\cdots\right)\nonumber\\
  I_{(1,1,-1)(1,-2)(3)}&=&x^{4b}\left(-18x^{10}-36x^{12}+198x^{14}+549x^{16}+\cdots\right)\nonumber\\
  I_{(1,1,-1)(1,-2)(2,1)}&=&x^{4b}\left(90x^8+45x^{10}-36x^{12}+\cdots\right)\nonumber\\
  I_{(1,1,-1)(1,-2)(1,1,1)}&=&x^{4b}\left(0x^{12}+\cdots\right)\nonumber\\
  I_{(1,1,-1)(1,-1,-1)(3)}&=&x^{4b}\left(36x^{14}+\cdots\right)\nonumber\\
  I_{(1,1,-1)(1,-1,-1)(2,1)}&=&x^{4b}\left(-180x^{12}+\cdots\right)\nonumber\\
  I_{(1,1,-1)(1,-1,-1)(1,1,1)}&=&x^{4b}\left(0x^{12}+\cdots\right)
\end{eqnarray}
Note the explicit appearance of the trial R-charge $b$. At generic $b$, this does not
cancel with any other term in the field theory index: for instance, the first term $9x^{4b+6}$
on the first line, which is larger than $x^{10}$, cannot combine with other terms listed above
and below. In particular, for the field theory index to match with the gravity index
including these sectors, $b$ has to be half an integer as the gravity spectrum of $\epsilon+j_3$
is all even. The fluxes in the last case are obtained by flipping the second and third fluxes.
We find that, up to the order that we wish to check, these sectors all give zero contributions
(which might mean that they are identically zero):
\begin{eqnarray}
  I_{(2,-1)(3)(1,-2)}&=&0x^{16b}x^{20}+\cdots,\
  I_{(2,-1)(2,1)(1,-2)}=0x^{16b}x^{20}+\cdots,\nonumber\\
  I_{(2,-1)(1,1,1)(1,-2)}&=&0x^{16b}x^{12}+\cdots,\
  I_{(2,-1)(3)(1,-1,-1)}=0x^{16b}x^{12}+\cdots\nonumber\\
  I_{(2,-1)(2,1)(1,-1,-1)}&=&0x^{16b}x^{12}+\cdots,\
  I_{(2,-1)(1,1,1)(1,-1,-1)}=0x^{16b}x^{12}+\cdots\nonumber\\
  I_{(1,1,-1)(3)(1,-2)}&=&0x^{16b}x^{12}+\cdots,\
  I_{(1,1,-1)(2,1)(1,-2)}=0x^{8b}x^{10}+\cdots\nonumber\\
  I_{(1,1,-1)(1,1,1)(1,-2)}&=&0x^{8b}x^{10}+\cdots,\
  I_{(1,1,-1)(3)(1,-1,-1)}=0x^{8b}x^{10}+\nonumber\\
  I_{(1,1,-1)(2,1)(1,-1,-1)}&=&0x^{8b}x^{10}+\cdots,\
  I_{(1,1,-1)(1,1,1)(1,-1,-1)}=0x^{8b}x^{10}+\cdots\ .
\end{eqnarray}
We want to compute the full index up to $\mathcal{O}(x^{14})$ and show that this
factorizes into the product of `positive-flux part' $I_{(P_+,P_-)\!=\!(2,0)}$ and
`negative-flux part' $I_{(P_+,P_-)\!=\!(0,1)}$ after summing over all the
saddle points with $P_+\!=\!2$, $P_-\!=\!1$.
Summing over $32$ saddle points with $P_+=2,P_-=1$, one obtains
\begin{eqnarray}\label{multi-2-1}
  I_{(P_+,P_-)=(2,1)}&=&\hat{I}_{(P_+,P_-)\!=\!(2,1)}+I^b_{(P_+,P_-)\!=\!(2,1)}\nonumber\\
  &\equiv&9x^2\left(63x^4-45x^8+360x^{10}-1251x^{12}+\cdots\right)+
  (405x^{10}-3240x^{12}+11259x^{14}+\cdots)\nonumber\\
  &&+x^{4b}\left(9x^6+18x^8-x^{10}+\cdots\right)\ ,
\end{eqnarray}
where $\hat{I}$ is the expression on the second line while $I^b$ is the expression on
the last line containing the trial R-charge $b$. The first parenthesis of the second line
is from the first six saddle points with factorized indices, the second term from the next
three with unfactorized indices, and the last line is from the next $12$ saddle points
whose leading energy levels depend on the undetermined trial R-charge $b$. It is easy to
find that the first two contributions show a nontrivial cancelation and yield
$\hat{I}=(9x^2)(63x^4)=I_{(P_+,P_-)\!=\!(2,0)}I_{(P_+,P_-)\!=\!(0,1)}$, and this agrees
with the graviton index consisting of positive charged particles with net charge $2$ and
negatively charged particles with charge $-1$,
\begin{equation}
  \hat{I}(x)=I^{(-1)}_{\rm sp}(x)\left(I^{(+2)}_{\rm sp}(x)+\frac{I^{(+1)}_{\rm sp}(x^2)+
  I^{(+1)}_{\rm sp}(x)^2}{2}\right)=(9x^2)(63x^4)\ .
\end{equation}
On the other hand, the $I^b$ contribution does not appear to correspond to anything
in the gravity dual. Especially, considering the allowed range $0<b<1$ from $a,b,c>0$,
the first term is or lower order than $\mathcal{O}(x^{10})$. As we have explained,
if these saddle points are artifacts of our localization calculations introduced by
unphysical saddle points `flowing in from infinite' \cite{Witten:1992xu}, the remaining
index completely agrees with the gravity index. This however requires a careful study,
similar to \cite{Kim:2010ac}.

At this stage, let us comment that we empirically find from the the above saddle points
a rule for the indices which do not contain $b$. Namely, in the first eight saddle points,
if one decomposes the positive and
negative fluxes into two, each of them separately satisfies the flux condition
(\ref{overall-decouple}) for the fluxes. Note that this restriction turns out to pick up the
factorizable saddle points only.

\section{Discussions}

We initiated the index computation of Chern-Simons matter theories
with $N=2,3$ theories. We find the perfect matchings between the
field theory index and the gravity index for $N=3$ theory, which
describes M2 brane probing $N^{010}/Z_k$ theory.  On the other hand
we find the several subtleties for $N=2$ models even with the
impressive matching with the gravity index. For $Q^{111}$ model,
which is realized as flavored ABJM model, the field theory index
gives the definite prediction for the gravity index, which  begs the
reexamination of the KK spectrum of $Q^{111}$ in the gravity side
available in the literature.

For $M^{32}$ model, the presence of odd number ($i=1,2,3$) fermions
in each bi-fundamental sector seems to pose a potential issue on
whether this theory is consistent or not. At least in the context of
dealing with QFT on $S^2\times S^1$ and partition function there, we
find that appropriately restricting the magnetic charges of gauge
fields in the overall $U(1)$'s of $U(N)$'s on $S^2$ makes the resulting
partition function well-defined. This is basically due to 
shifting of Chern-Simons level for $U(1)$ factors (among others) out
of $U(N)$ due to fermion determinant. It would be interesting to work out
the structure of the potential mixed Chern-Simons terms.
 The comparison with the known gravity index is amazingly
impressive but we face handling with the new saddle points. It could
be important to understand this problem to understand better
AdS4/CFT3 correspondence for vast classes of $N=2$ theories. Even
though we do not report our computation of the index for $Q^{111}$
model suggested at \cite{Franco:2008um}, they are suffering from the
same problems as our $M^{32}$ model has, i.e., potential anomaly
issues due to chiral fermions and new saddle points. Now these
problems are not appearing in the $Q^{111}$ model \cite{Jafferis:2009th,Benini}
we considered, we hope to construct new $M^{32}$ model free from such subtleties,
yet giving the impressive matching with the gravity index, which our
current $M^{32}$ model exhibits. We hope to report our progress in
this direction in near future. However, we emphasize that final
verdict on the current $M^{32}$ model has not been made yet.

\vskip 0.5cm  \hspace*{-0.8cm}
{\bf\large Acknowledgements}
\vskip 0.2cm

\hspace*{-0.75cm} We are grateful to Francesco Benini, Daniel Jafferis, Hee-Cheol Kim,
Nakwoo Kim, Igor Klebanov, Silviu Pufu and Benjamin Safdi for helpful discussions. This work
is supported by the BK21 program of the Ministry of Education, Science and Technology (DG, SK),
KOSEF Grant R01-2008-000-20370-0 (JP), the National Research Foundation of Korea (NRF) Grants
No. 2009-0085995 (JP), 2010-0007512 (SC, SK) and 2005-0049409 through the Center for
Quantum Spacetime (CQUeST) of Sogang University (JP). JP appreciates APCTP for
its stimulating environment for research. SK thanks Princeton Center
for Theoretical Science (PCTS) for hospitality during his visit, where this work
was completed.

\end{document}